\documentclass[a4paper,11pt]{JHEP3}
\pdfoutput=1

\usepackage{latexsym}
\usepackage{amsmath}
\usepackage{amssymb}
\usepackage{mathrsfs}
\usepackage{dsfont}
\usepackage{slashed}
\usepackage{stmaryrd}
\usepackage{graphicx}
\usepackage{accents}

\newcommand{\hal}{{\textstyle\frac{1}{2}}}
\newcommand{\ca}[1]{{\mathcal{#1}}}
\newcommand{\bs}[1]{{\boldsymbol{#1}}}
\newcommand{\unit}{\mathds{1}}  
\newcommand{\fr}[2]{{\textstyle{\frac{#1}{#2}}}}

\newcommand{\del}{\partial}
\newcommand{\vep}{\varepsilon}
\newcommand{\vphi}{\varphi}

\newcommand{\ww}{\underline}

\newcommand{\Nf}{N_f}
\newcommand{\sN}{{\mathcal{N}}}
\newcommand{\aux}[1]{{\mathsf{#1}}}
\newcommand{\mop}{\mathcal{M}}

\newlength{\dhatheight}
\newcommand{\hhat}[1]{%
    \settoheight{\dhatheight}{\ensuremath{\hat{#1}}}%
    \addtolength{\dhatheight}{-0.35ex}%
    \hat{\vphantom{\rule{1pt}{\dhatheight}}%
    \smash{\hat{#1}}}}

\author{Robert Wimmer\\
Universit\'e de Lyon, Laboratoire de Physique, UMR 5672, CNRS, \\
\'Ecole Normale Sup\'erieure de Lyon,\\
46, all\'ee d'Italie, F-69364 Lyon cedex 07, France\\
{\tt robert.wimmer@ens-lyon.fr}
}

\title{An index for confined monopoles}

\dedicated{Dedicated to the memory of Francis A. Dolan}

\abstract{We compute the index and associated spectral density for fluctuation operators
which are defined via the Lagrangian of $\sN =2$ SQCD in the background of non-abelian confined multimonopoles.
To this end we generalize the standard index calculations of Callias and Weinberg to the case of 
asymptotically nontrivial backgrounds. The resulting index is determined by topological 
charges. We conjecture that this index counts one quarter of the dimension of the moduli 
space of confined multimonopoles.}

\begin{document}
\newpage
   
\section{Introduction}

From their discovery on  \cite{Auzzi:2003fs,Tong:2003pz}, 
non-abelian confined  monopoles and vortices received 
considerable attention. The reason for this is that these objects may allow 
the description of non-abelian confinement as an electric-magnetic dual 
Meissner effect, as it was envisioned by 't Hooft \cite{Hooft:1975pu}
and Mandelstam \cite{Mandelstam:1974vf}. We refer to \cite{Konishi:2007dn, 
Tong:2008qd,Shifman:2009zz} for recent reviews of these developments. 

In \cite{Burke:2011mw}, among other things, the full perturbative quantum energies and a central charge
anomaly for confined multimonopoles in $\sN =2$ SQCD have been computed. 
Central to this quantum computation is the spectral density of certain operators, 
which can be obtained from an index theorem.  The present investigation is devoted to the derivation of 
these quantities. 
The operators in question are the fluctuation operators obtained from the expansion of the 
Lagrangian around the classical background fields that satisfy the $\fr{1}{4}$ BPS
equations which describe confined monopoles (and many other field configurations). 
These operators, describing the fluctuations of the full
second order equations of motion, differ from the fluctuation operators that 
describe the fluctuations of the first order $\fr{1}{4}$ BPS  equations. We will discuss 
this in detail. 

\begin{figure}[tp]
 \label{fig}
  \centering
  \includegraphics[width=15cm]{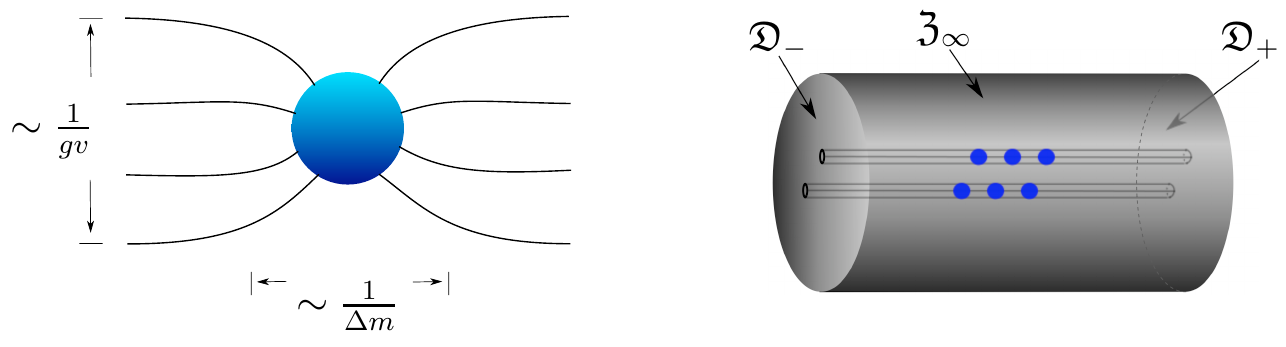}
  \caption{The left figure shows a confined monopole in the close up with characteristic
    scales for the monopole size, displaying approximate Coulomb monopole behavior, 
    and the size of the confining vortices. The right figure depicts the spatial boundary, a cylinder at infinity,
    where the asymptotic behavior of the (multiple) confined monopoles is specified (see main text). The cylinder 
    at infinity consists of the two \emph{discs} $\mathfrak{D}_{\pm}$ at $x^3\rightarrow\pm \infty$ and the 
    \emph{cylinder wall} $\mathfrak{Z}_{\infty}$ at $r\rightarrow\infty$, with $r$ being the  
    cylindrical radial coordinate.}
\end{figure}

The index has been computed for many different topologically non-trivial backgrounds in different models, 
in the given context for example for vortices and domain walls \cite{Hanany:2003hp,Sakai:2005sp}, 
but not for confined monopoles. The difference in the case of confined monopoles is that the
usual techniques developed in \cite{Callias:1977kg,Weinberg:1979ma,Weinberg:1981eu} cannot 
be  applied directly.  Confined multimonopoles depend on all 
three spatial coordinates, but the nontrivial field dependence is essentially 
concentrated in flux tubes, i.e.\ the vortices that confine the monopole and emanate from 
it, see figure \ref{fig}. Consequently, the background fields of the fluctuation operators do not 
fall off asymptotically, or terminate in a vacuum configuration. We describe in the 
following sections how to resolve this problem and develop a strategy that generalizes
the established methods  \cite{Callias:1977kg,Weinberg:1979ma,Weinberg:1981eu} 
of the (open space) Callias index theorem
to the case of asymptotically nontrivial backgrounds.
 
Following \cite{Callias:1977kg}, the index of such operators can be written as the sum of an anomaly 
and a boundary or surface term. Both contributions can be conveniently computed as appropriate 
limits, however, for this to be true for the surface term the background fields 
have to be asymptotically trivial. 
The latter does not apply to the field configurations considered in this paper.
For the concrete case of $\fr{1}{4}$ BPS confined monopoles, the principle form of the field 
configurations is depicted in figure \ref{fig}. The given geometrical setting defines the 
surface term on a boundary of the form of a cylinder at infinity. Hereby, one has especially
at the discs $\mathfrak{D}_{\pm}$ highly nontrivial field configurations, notably multiple vortices which 
are not even known analytically. We will show how to reformulate these surface contributions 
in the form of an index, though of a generalized form. Its computation can again be reduced to an anomaly
and a surface term on the boundary $\del\mathfrak{D}_{\pm}$ of the discs. It turns out that only the anomaly on
the discs gives a non-vanishing contribution, which however depends on the IR-regulator and 
thus leads to a non-vanishing spectral density. Based on some explicit examples we conjecture that 
the resulting index counts one quarter of the real dimension of the moduli space for general 
confined multimonopoles.

The paper is organized as follows: In section 2 we introduce the necessary details of the model, 
the BPS equations and the associated fluctuation operators. In section 3 we briefly review the 
conventional techniques, introduce the necessary generalizations and compute the index as a function 
of the background fields. We also relate the resulting index to topological charges and formulate 
a conjecture regarding its relation to the dimension of the moduli space of
confined multimonopoles. Finally we discuss the relation between the fluctuation operators 
obtained from the Lagrangian and those obtained directly from the BPS equations. In section 4 
we summarize our results and give conclusions. In particular, we give an outline for the general 
recipe to compute the index in the presence of asymptotically nontrivial backgrounds. 
In the appendices we give several details for
the proofs of the statements in the main text.

\section{Confined Monopole   BPS Equations}

\subsection{Model and vacua}

The full Lagrangian for $\sN=2$ SQCD with generalized FI-terms and the conventions that we use  
are given in \cite{Burke:2011mw}. Here we need only the bosonic part of the Lagrangian 
which is given by\footnote{For the 
metric we use the east coast convention $\eta_{\mu\nu}=(-,+,+,+)$ and 
generally summation over repeated flavor indices $i=1,\ldots,N_f$ is implied. The covariant derivative is defined 
as $D_\mu=\partial_\mu-i A^R_\mu$ for fields in the representation $R$. The field strength is
given by $F_{\mu\nu}=\del_\mu A_\nu-\del_\nu A_\mu-i[A_\mu,A_\nu]$. The positions of 
$SU(2)_R$ indices $I$  are changed by complex conjugation.}
\begin{align}
\label{lsur}
  \ca{L}_{\mathrm{bos}} =&\  -\fr{2}{g^2}\,\mathrm{Tr}\,\{\, \fr{1}{4} F_{\mu\nu}^2 + |D_\mu \phi|^2
          + \fr{1}{2}[\phi,\phi^{\dagger}]{}^2 + \fr{1}{2}\,{\vec{\aux{D}}}^2\, \} \nonumber\\[5pt]
     &\hspace{2cm}  -  |D_\mu S_{i I}|^2  -\bar S^I_i\{\phi^\dagger+\bar m_i,\phi+m_i\}S_{iI} \, .
\end{align}
The complex scalar field $\phi$ transforms in the adjoint representation of the gauge 
group $U(N)$, whereas the $N_f$  $SU(2)_R$ doublets $S_{iI}$ are scalars that 
transform in the 
fundamental representation of $U(N)$ and the flavor symmetry $SU(N_f)$. The latter one is
explicitly broken to a subgroup by the masses. The bracket in the last term indicates the anti-commutator 
of the given matrices. We are interested in static $\frac{1}{4}$ BPS solutions and therefore 
we choose the masses to be real (and ordered): $\bar m_i = m_i$ and $m_1 \leq \ldots \leq m_{N_f}$.
See \cite{Burke:2011mw} for a detailed discussion.
The auxiliary field 
$\vec{\aux{D}}$ is an  $SU(2)_R$ vector which is in the adjoint representation of the gauge group.
We will keep the auxiliary field for notational convenience but it is understood
to be on-shell, i.e.
\begin{equation}
  \label{d}
  \vec{\aux D}= \frac{g^2}{2}\, (\vec{\tau}_I{}^J S_{i\,J}\otimes \bar S^I_i - \vec{\xi}\ \unit )\ ,
\end{equation}
where the tensor product notation is defined w.r.t.\ the gauge group structure carried by the fields.

The $SU(2)_R$ triplet $\vec{\xi}$ of FI-terms explicitly breaks the $SU(2)_R$ symmetry 
and can therefore be rotated to a convenient choice \cite{Shifman:2009zz}. 
For the rest of the article we 
will assume that it points in the positive three-direction which we parametrize as
\begin{equation}
  \label{xi}
 \vec{\xi} =  (0,0,v^2)^t \, ,
\end{equation}
with $v$ being a positive constant.

For the generators of the gauge group $U(N)$ we have the following conventions: 
The hermitian generators are $\{T^A\}=\{T^0,T^a\}$, where $\{T^a\}$ forms an $\mathfrak{su}(N)$ 
algebra, and  satisfy
\begin{equation}
  \label{eq:un}
  [T^a,T^b]=i f^{abc}\, T^c\ ,\ \ T^0=\fr{1}{\sqrt{2 N}}\unit\ , 
     \ \ \mathrm{Tr}\{T^A T^B\}=\fr{1}{2}\,\delta^{AB}\ ,
\end{equation}
where $f^{abc}$ are the real and totally antisymmetric $\mathfrak{su}(N)$ structure constants.\\

\noindent
{\bf Color-Flavor locked vacua.} The Lagrangian (\ref{lsur}) provides a particular set of vacua 
which preserve a diagonal subgroup of the gauge group $U(N)$ times the unbroken flavor 
group $H_F\subset SU(N) \hookrightarrow SU(N_f)$.\footnote{The focus is here on the part 
of the unbroken flavor group that acts nontrivial on the to be considered background fields which 
carry flavor index $i=1,\ldots,N$.} Up to gauge transformations the vacuum is specified by the 
following vacuum values of the scalars:
\begin{equation}
 \label{vac1}
 \phi_0 = - \mathrm{diag}\,(m_1,\ldots,m_N)   \ ,  \quad\quad
  [S_1^{\mathrm{vac}}]^{n}{}_i = v\,\delta^n{}_i \, ,
\end{equation}
where the fundamental scalar $S_1^{\mathrm{vac}}$ is written as a $N\times N_f$ matrix,
whose entries are zero for $i > N$. These vacuum scalars preserve certain subgroups
of the original symmetry, i.e.
\begin{equation}
  \label{vac2}
   H_C\, \phi_0\, H^{-1}_C = \phi_0\ , \quad\quad     
   H_{C+F} \,S_1^{\mathrm{vac}} \,H^{-1}_{C+F} = S_1^{\mathrm{vac}}\ .
\end{equation}
The transformation $H_{C+F}$ acts from the left as a global $U(N)$ gauge transformation,
and from the right as an $H_F$ flavor transformation. Hence, one has  
$H_{C+F} \subset \mathrm{diag}\, (\, U(N)\times SU(N_f)\,)$ and such vacua are
called the color-flavor locked phase. 

The associated symmetry breaking pattern is given by
\begin{equation}
  \label{vac4}
  U(N)\times SU(\Nf)  \overset{m}{\longrightarrow} U_C(1)\times H_C \times H_F 
  \overset{v}{\longrightarrow} H_{C+F} \, ,
\end{equation}
where we assumed that all masses are of the same scale and that $m \gg v$. 
If the first $N$ masses form  
$q$ groups of $n_r$ degenerate masses the surviving symmetry group is given by \cite{Nitta:2010nd},
\begin{equation}
  \label{HCF}
  H_{C+F}= S(\, U(n_1)\times \ldots \times U(n_q)\,)\, ,
\end{equation}
with $\sum_r n_r =N$. It supports monopoles with typical size $1/\Delta m$, the 
inverse mass difference, that
are confined by flux tubes of width $1/gv$, see figure \ref{fig}.

\subsection{ $\frac{1}{4}$ BPS equations and asymptotics}\label{secasym} 

We denote  
the classical background fields that satisfy the BPS equations given below by 
\begin{align}
  \label{bkg}
   A_k^{\mathrm{bkg}} = \ca{A}_k\ ,\quad 
   \phi_{\,\mathrm{bkg}}=\vphi = \vphi^\dagger \ , \quad
    S_{i1}^{\mathrm{bkg}} = \Sigma_{i} \  , \quad 
    \aux{D}^3_{\mathrm{bkg}}= \mathsf{D} \, , 
\end{align}
where the index $k=1,2,3$ runs over the three spatial directions. 
The covariant derivative w.r.t.\ the gauge field $\ca{A}_k$ will be denoted by  $\ca{D}_k$. 
All other fields are assumed to be zero for the classical solutions of interest, in particular one has 
$\Sigma_{i\, >\, N}=0$. This ansatz  implies also
that classically the adjoint scalar $\phi$ is hermitian.

With the particular choice (\ref{xi}) for  $\vec{\xi}$, the first description of confined monopoles in the 
given context and the associated novel BPS equations were given in \cite{Tong:2003pz}, derived using the 
Bogomolnyi trick. It was shown that the energy density is (locally) minimized if the following 
first order equations are satisfied:
\begin{align}
  \label{BPS}
  &\ca{B}_k -\sqrt{2}\, \ca{D}_k\vphi + \delta_{k\,3}\, \aux{D} = 0 \quad\quad
  \textrm{with:} \ \ \aux{D}=\fr{g^2}{2}\,(\, \Sigma_i\otimes\bar{\Sigma}_i -v^2\,) \nonumber\\[3pt]
 &\ca{D}_z \Sigma_i = 0 \  , \ \ \ca{D}_w \Sigma_i = 0\, ,
\end{align}
where we have introduced complex coordinates such that $\ca{D}_z = \ca{D}_1 + i \ca{D}_2$ and
$\ca{D}_w = \ca{D}_3 + i \ca{D}_4$, with $ \ca{D}_4 = i\,\sqrt{2} (\vphi +m_i)$ when acting on 
the $i$'th fundamental scalar $\Sigma_i$ (in the adjoint representation the 
mass $m_i$ drops out). The chromo-magnetic field is defined as $\ca{B}_k =\fr{1}{2} \vep_{kij}\,\ca{F}_{ij}$.

These equations a priori look overdetermined, 
but as was noted in \cite{Isozumi:2004vg} the equations for $k=1,2$ are identical to the integrability 
condition of the last one. This novel set of equations offers a plethora of non-trivial field 
configurations carrying various topological charges 
\cite{Sakai:2005sp}. For us the main focus lies with configurations that describe confined monopoles. 
In \cite{Shifman:2004dr} an approximate solution for a single
confined monopole was given for the case of  $U(1)\times SU(2)$ gauge group
and $\Nf=2$ (though for a different choice of $\vec{\xi}$ ). The relation of these equations to the 
supersymmetry algebra and the associated (tensorial) central charges were discussed in detail in 
 \cite{Burke:2011mw}.
Equations with less supersymmetry for intersecting vortices were derived in \cite{Eto:2005sw}.\\

\noindent
{\bf Classical Asymptotics.}\label{asymbeh} 
We consider those solutions of 
(\ref{BPS}) that describe  multiple confined monopoles of the form as depicted  
in figure \ref{fig},
such that the field configurations have an axial orientation in the $x^3$-direction. The main 
input for the index calculation will be the asymptotic behavior, i.e. the topology, of 
these solutions.

The axial orientation of the field configurations implies that the asymptotic boundary has the form of an 
infinite cylinder, see figure \ref{fig}, and accordingly we have to specify the asymptotic behavior: 
i.) At $x^3 \rightarrow \pm\infty$ the boundary is given by the infinite discs $\mathfrak{D}_{\pm}$
at which the fields behave like multiple vortices, though in general different vortices at 
$\mathfrak{D}_{+}$ and $\mathfrak{D}_{-}$. ii.) For $r \rightarrow \infty$, with $r$ being the 
radial cylindrical coordinate, the boundary is given by the cylinder wall  $\mathfrak{Z}_\infty$
 at  infinity.  The flux is confined in the vortices
which are infinitely far away from the cylinder wall $\mathfrak{Z}_\infty$ and therefore vanishes
exponentially with correlation length proportional to $gv$, see (\ref{BPS}). Hence, at the 
cylinder wall one has asymptotic vortex behavior, with winding in the Higgs fields and the 
long-ranged gauge fields. Due to the monopoles this winding depends on $x^3$, but this dependence
is exponentially located at the monopoles with the characteristic length given by the associated mass
difference $|\Delta m|$. Concretely, the asymptotic field behavior is as follows:
\\

\noindent
{\underline{Cylinder Wall $\mathfrak{Z}_\infty$:}}  The confinement of the monopoles/flux by vortices implies   
$\ca{B}_k \approx 0 \approx \ca{A}_r$, where $\approx$ means equal up to
exponentially suppressed terms $\sim e^{-gvr}$. The Higgs fields $\Sigma_i$ 
approach their vacuum values (up to winding) exponentially fast, i.e.\ with the spatial angular 
coordinate $\theta$ one has,
\begin{equation}
  \label{asym1}
  \Sigma_i\approx U S_{1i}^{\mathrm{vac}} \quad , \quad U = e^{i\theta (w+w_{C+F})} \,  .
\end{equation}
The part of $U$ which lies in the unbroken color-flavor symmetry $H_{C+F}$ (\ref{HCF}) is generated
by $w_{C+F}(x^3)$ and has a kink-like $x^3$-dependence,  localized at the monopoles. 
For $x^3\rightarrow \pm\infty$ the matrix $U$ is in the 
Cartan subgroup, see \cite{Shifman:2004dr}
for an explicit example. % The total vortex number is determined by $w$ since $\mathrm{Tr}\,w_{C+F}=0$. 
The asymptotic form of the BPS equations (\ref{BPS}) determines the residual fields to be of the form,
\begin{equation}
  \label{asym2}
  \ca{A}_3 \approx i U\del_3\, U^{-1}\ \ ,\ \ \ca{A}_{\theta}\approx i U\del_\theta\, U^{-1}\ \ , \ \ 
  \vphi \approx U\phi_0U^{-1}\ .
\end{equation}
In addition to the  BPS equations (\ref{BPS}) one finds that asymptotically 
 $\ca{D}_{\bar z}\Sigma_i \approx \ca{D}_{\bar w}\Sigma_i \approx \aux{D}\approx 0$. 
%What is indicated here for the asymptotic behaviour, that the $x^3$ dependence enters through 
%moduli $w_{C+F}$ due to the 
\\

\noindent
{\underline{Discs $\mathfrak{D}_\pm$:}} On the discs at infinity the fields approach
 pure vortex behavior  exponentially fast, with suppressed corrections $\sim e^{-|\Delta m \Delta x^3|}$ 
($\Delta x^3$ being the distance to the monopoles). 
Therefore one has $\ca{B}_1\approx\ca{B}_2\approx\ca{A}_3\approx\ca{A}_r\approx 0$ and $\vphi\approx \phi_0$. 
The nontrivial fields $\ca{B}_3, \ca{A}_\theta$ in general take different values at the two  discs 
at infinity, but not the abelian $U(1)$ part. In particular one has, 
\begin{equation}
  \label{asym3}
  \Sigma_i{}^m|_{\mathfrak{D}_\pm} \approx 
  \mathrm{diag} (\sigma_1^{\pm},\ldots, \sigma_N^{\pm})\ \ , \ \ 
   \ca{A}_{\theta}|_{\mathfrak{D}_\pm} = \ca{A}^{(\pm)}_{\theta} \ \ , \ \ 
   \mathrm{Tr}\, \ca{A}_{\theta}|_{\mathfrak{D}_{+}} =  \mathrm{Tr}\, \ca{A}_{\theta}|_{\mathfrak{D}_{-}}\, ,
\end{equation}
and analogously for $\ca{B}_3|_{\mathfrak{D}_{\pm}}$. The BPS equations (\ref{BPS}) imply then 
$[\ca{A}_{\alpha=1,2},\, \phi_0] \approx 0$ so that $[\ca{B}_{3},\, \phi_0]\approx 0$.  In addition 
one finds $\ca{D}_{\bar w} \Sigma_i \approx 0$. 
\\

\subsection{Fluctuation operators}
The main focus of the investigation in \cite{Burke:2011mw} was on the quantum properties
of confined monopoles. For this purpose the full $\sN=2$ SQCD Lagrangian had to be expanded to 
second order in the quantum fields in a background that satisfies the BPS equations (\ref{BPS}),
i.e.
\begin{align}
  \label{fluc}
  A_{k} &\ =\ \ca{A}_{k} + a_{k}\, ,&  \sqrt{2}\phi &\ =\  \sqrt{2}\, \vphi - a_4 + i a_5
   \nonumber\\
  S_{i1} &\ =\ \Sigma_{i 1} + s_{i1} \, ,&    S_{i2} &\ =\ s_{i2}\, .
\end{align} 
With a convenient gauge fixing term the Lagrangian quadratic in the 
quantum fields can be written in the form
\begin{align}
  \label{l2}
  \ca{L}_{\mathrm{bos}}^{(2)} =&\ 
     - \mathrm{Tr}\,\{ W^\dagger \, \big{(}\, \del_{0}^2\, +\,  L L^\dagger\,\big{)}\,W\,\} 
                + \ldots\ , 
\end{align}
where the four-component fluctuation field $W$ is given by,
\begin{equation}
  \label{W}
  W = [\ \fr{1}{g}\, a_w\, ,\ \fr{1}{g}\, a_z\, ,\ s_{i\,1}\,,\ s_{i\,2}\ ]^T \, ,
\end{equation}
with the complex coordinates $z,w$ defined as in (\ref{BPS}). We emphasize that here
also the trivial fluctuations $s_{1\, i>N_f}$ and $s_{i2}$ around the vanishing backgrounds 
$\Sigma_{i>N_f}=S_{i2}=0$ 
are included. This will be discussed below.
  
The ellipsis stands for terms of a similar structure. These are ghost fields and the trivial $a_5$ and $a_0$ 
fluctuations, which are governed by the one-one matrix component of $L^\dagger L$. 
There are no zero modes from these fields, as we will see. 
The full supersymmetric Lagrangian also includes fermionic fluctuations governed by operators 
$L$ and $L^\dagger$, which contribute zero modes for the classical e.o.m., 
and are in fact the reason to organize the bosonic fluctuations according to (\ref{l2}). 
We refer to  \cite{Burke:2011mw}
for more details, which are not relevant for the present considerations. 

The objects of
main interest are the fluctuation operators $L,L^\dagger$ which depend on the 
classical background fields defined by the BPS equations (\ref{BPS}). Their detailed form 
is
\begin{equation}
  \label{LLd}
  L= \left[
     \begin{array}{cc}
     \slashed{\ca{D}}^{\mathrm{a}}  &  - ig\, \bar{\Sigma}_i^{\,\mathrm{r}} \\[3pt]
     ig\, \Sigma_i^{\,\mathrm{r}}   &  \bar{\slashed{\ca{D}}}{}^{\mathrm{f}}
     \end{array} \right] 
      \ , \quad \quad 
  L^\dagger= \left[
     \begin{array}{cc}
     -\bar{\slashed{\ca{D}}}^{\mathrm{a}}  &  - ig\, \bar{\Sigma}_i^{\,\mathrm{r}} \\[3pt]
     ig\, \Sigma_i^{\,\mathrm{r}}   &  - \slashed{\ca{D}}{}^{\mathrm{f}}
     \end{array} \right] \, , 
\end{equation}
where we introduced Euclidean quaternions and the abbreviations,
\begin{align}
  \label{euk}
  \sigma^{\ww k} = \ (\,\sigma^k, i\unit_2 \,)\, , \  
     \slashed{\ca{D}} = \sigma^{\ww k}\, \ca{D}_{\ww k} \  \  \ \ \
   \textrm{and}\ \ \ \ \
  \bar\sigma^{\ww k} = \ (\,\sigma^k, - i\unit_2 \,)\, , \  
      \bar{\slashed{\ca{D}}} = \bar\sigma^{\ww k}\, \ca{D}_{\ww k} \, .
\end{align}
The index $\ww k$ runs now over four values and $\ca{D}_4$ was introduced below (\ref{BPS}).
The superscripts $\mathrm{a}, \mathrm{f}$  indicate the adjoint and fundamental representation, whereas
the superscript $\mathrm{r}$ indicates action from the right. Explicitly, the right action on an adjoint
field $X$ is just matrix multiplication, $\Sigma_i^{\, \mathrm{r}} \cdot X := X\Sigma_i$, and on a 
fundamental field $y_i$ it is tensor multiplication
$\bar{\Sigma}_i^{\, \mathrm{r}} \cdot y_i := y_i\otimes\bar\Sigma_{i}$.  
As usual, summation over repeated flavor indices  is  implied. 

The operators act in the space of the direct sum of adjoint and fundamental fields, for example 
$L:\left[{X\atop y_i}\right] \rightarrow \left[{X'\atop y'_i}\right]$. It is with regard to the 
natural scalar  product in this space that the adjoint operator $L^\dagger$ is 
the hermitian conjugate of $L$, and vice versa.

We want to emphasize that the fluctuation operators (\ref{LLd}) originate from the quadratic 
Lagrangian in the BPS background and thus describe the fluctuations w.r.t.\ the full, second
order field equations. This is in contrast to the fluctuations of the 
$\frac{1}{4}$ BPS equations (\ref{BPS}). We will discuss this subtlety in more detail below. 

In order to compute quantum corrections the quantity of prime interest is the difference in 
the spectral densities of the operators $LL^\dagger$ and  $L^\dagger L$. We will call this quantity 
henceforth ``spectral density'' $\Delta \rho$. This spectral density can be conveniently
extracted from an index theorem. The advantage of the index theorem calculation is 
that usually only the asymptotic behavior, i.e. the topological properties of the 
classical background fields, have to be known. In the case of confined monopoles this is not a 
straight forward issue as we will discuss now.

\section{Index Theorem}

The index of an operator $L\,$, $\mathrm{Ind}(L) = n_{L^\dagger L}^0 - n_{L L^\dagger}^0\,$, 
counts the difference in the number of zero modes 
of\footnote{A simple argument for the norm of zero modes also shows that 
$L^\dagger L \psi_0=0 \Leftrightarrow L\psi_0=0$ and $LL^\dagger \psi_0=0 \Leftrightarrow L^\dagger\psi_0=0$.}
$L$ and $L^\dagger$, and can be obtained
from an IR-regulated expression:  
\begin{equation}
  \label{ind1}
  \ca{I}(M):= \mathrm{T}_{\mathrm{R}}\,\Big{\{} \frac{M^2}{L^\dagger L+M^2}-\frac{M^2}{LL^\dagger+M^2}\,\Big{\}}
   \ \ \  \Rightarrow \ \ \ 
   \mathrm{Ind}(L) = \lim_{M\rightarrow 0} \ca{I}(M)\ ,
\end{equation}
where the modified symbol for the trace indicates that the trace is taken now also over the functional 
Hilbert space. Of particular interest for us is the application to non-compact spaces, 
as it was developed in \cite{Callias:1977kg,Weinberg:1979ma,Weinberg:1981eu}. 

The spectral density $\Delta \rho$ can be extracted by a Laplace transformation 
from the continuum contribution to the index function $\ca{I}(M)$:
\begin{equation}
  \label{ind2}
  \ca{I}^{\mathrm{cont}}(M) := \ca{I}(M)-\ca{I}(0) = 
 \int_{\omega}\, \frac{- M^2}{\omega^2+M^2}\ \Delta \rho\, ,
\end{equation}
where the signs are chosen such that they match the conventions of \cite{Burke:2011mw}. 
The measure for the integration $\int_\omega$ over the continuum-mode eigen-values
$\omega$ is defined via the l.h.s.

The usual technique to compute the index
is to transform $\ca{I}(M^2)$ into an anomaly term, which can be evaluated for $M\rightarrow\infty$, 
and a surface term,  which can be conveniently computed if the fields are asymptotically trivial.
The index is therefore
determined by the topological properties of the background fields. 
As discussed above, the situation for confined monopoles is rather different.  
In the following we will introduce the necessary generalizations of index theorem calculations and 
develop techniques for the   case of confined monopoles considered here.\\

\noindent
The BPS equations (\ref{BPS}) imply a certain structure for the fluctuation operators and thus
the index function $\ca{I}(M)$ given in (\ref{ind1}). The 
basic input for the index calculation are not the operators $L, L^\dagger$ itself but their
products, which are of the form
\begin{align}
  \label{LL}
   L^\dagger L=& \left[
     \begin{array}{cc}
     - \bar{\slashed{\ca{D}}}\slashed{\ca{D}}^{\mathrm{a}} +g^2 (\Sigma_k\otimes\bar\Sigma_k)^{\mathrm{r}}  &   
      - ig\,(\bar{\slashed{\ca{D}}} \bar{\Sigma}_j)^{\,\mathrm{r}} \\[3.5pt]
     ig\, (\slashed{\ca{D}}\Sigma_i)^{\,\mathrm{r}}   &  
       -\delta_{ij}\slashed{\ca{D}}\bar{\slashed{\ca{D}}}{}^{\mathrm{f}} + g^2\,\bar\Sigma_j\Sigma_i
     \end{array} \right] 
      \ , \nonumber\\[8pt]
   L L^\dagger=& \left[
     \begin{array}{cc}
     - \slashed{\ca{D}}\bar{\slashed{\ca{D}}}^{\mathrm{a}} +g^2 (\Sigma_k\otimes\bar\Sigma_k)^{\mathrm{r}}  &   
      ig\,(\slashed{\ca{D}} \bar{\Sigma}_j)^{\,\mathrm{r}} \\[3.5pt]
     -ig\, (\bar{\slashed{\ca{D}}}\Sigma_i)^{\,\mathrm{r}}   &  
       -\delta_{ij}\bar{\slashed{\ca{D}}}\slashed{\ca{D}}{}^{\mathrm{f}} + g^2\,\bar\Sigma_j\Sigma_i
     \end{array} \right]\ . 
\end{align}
They also have a nontrivial matrix structure in flavor space and act by 
$\left[{X\atop y_j}\right] \rightarrow \left[{X'\atop y'_i}\right]$.
The $\fr{1}{4}$-BPS equations (\ref{BPS}) imply the following relations for the building blocks of 
these operators:
\begin{align}
  \label{ddbps}
  &\bar{\slashed{\ca{D}}}\slashed{\ca{D}} = \ca{D}_{\ww k}^2 - \sigma^3\,  \aux{D}  \ \ , \ \ 
    \slashed{\ca{D}}\bar{\slashed{\ca{D}}} = 
   \ca{D}_{\ww k}^2 +\sigma^k\, (\, 2\,\ca{B}_k +\delta_{k,3}\aux{D}\,) \ , \nonumber\\[5pt]
  &\slashed{\ca{D}} \Sigma_i = 
   \begin{pmatrix} 0 & \ca{D}_{\bar z}\\ 0 & -\ca{D}_{\bar w} \end{pmatrix}\Sigma_i\, ,  
   \ \
    \slashed{\ca{D}}\bar\Sigma_i =
   \begin{pmatrix} \ca{D}_{ w} & 0\\ \ca{D}_{ z} & 0 \end{pmatrix}\bar\Sigma_i\ ,
\end{align}
With 
$\bar{\slashed{\ca{D}}}$ being
the hermitian conjugate matrix, one obtains the remaining combinations for the expressions in the second 
line of (\ref{ddbps}).

Contrary to the usual situation for $\fr{1}{2}$-BPS backgrounds 
neither of the fluctuation operators $L, L^\dagger$ (\ref{LLd})
is necessarily positive definite. This can be seen from the  (positive)
norm of these operators:\footnote{We are interested in possible zero modes here. Therefore 
we can safely neglect surface terms which do not contribute for such normalizable and 
localized discrete states. Non-normalizable zero modes that lead to many subtleties for non-abelian
monopoles \cite{Bais:1997qy} do not alter the following conclusions.}    
\begin{align}
  \label{Lnorm}
   \left\| L\,  {\textstyle{\left[{X\atop y_i}\right]}}  \right\|^2 \ = \ & 
      \left\| \slashed{\ca{D}} X \right\|^2 + \big{\|} \bar{\slashed{\ca{D}}}\, y_i \big{\|}^2
      +g^2 \big{(} \left\|y_i\otimes \bar\Sigma_i \right\|^2 +  \left\| X \Sigma_i \right\|^2\big{)}
      \nonumber\\[3pt]
      &\hspace{2.5cm}\ -ig\, \big{(}\, \bar{y}_{i\,m}\, \slashed{\ca{D}} \Sigma_i^n\, X^m{}_n 
                - X^{\dagger\,m}{}_{n}  \bar{\slashed{\ca{D}}} \bar\Sigma_{i\,m}\, y_i^n \,\big{)}\, ,
   \nonumber\\[6pt]
\left\| L^\dagger\,  {\textstyle{\left[{X\atop y_i}\right]}}  \right\|^2 \ = \ &
      \left\| \bar{\slashed{\ca{D}}} X \right\|^2 + \big{\|} \slashed{\ca{D}}\, y_i \big{\|}^2
      +g^2 \big{(} \left\|y_i\otimes \bar\Sigma_i \right\|^2 +  \left\| X \Sigma_i \right\|^2\big{)}
      \nonumber\\[3pt]
     &\hspace{2.5cm}\ -ig\, \big{(}\, X^{\dagger\,m}{}_{n}  \slashed{\ca{D}} \bar\Sigma_{i\,m}\, y_i^n 
              - \bar{y}_{i\,m}\, \bar{\slashed{\ca{D}}} \Sigma_i^n\, X^m{}_n\,\big{)}\, ,
\end{align}
where the second line in both equations give the cross terms of the norm square 
which are not necessarily positive (we 
indicated the gauge group indices $m,n$ explicitly). Usually for one of the
two operators $L, L^\dagger$ this cross term vanishes. In that case the positive definite 
non-derivative terms imply that the l.h.s.\ vanishes only if the state $[X,y_i]^t$ itself is 
zero. Therefore the respective operator has no nontrivial zero mode. In the case at hand
one can see from (\ref{ddbps}) that neither for $L$ nor for $L^\dagger$ 
the non-positive cross terms vanish and therefore both operators in general will have 
non-trivial zero modes.

\subsection{Principle structure. Anomaly vs.\ anomaly}\label{princip}

Before actually calculating the index function $\ca{I}(M)$ given 
in (\ref{LLd}), we discuss the well known 
basic structure behind such calculations
\cite{Callias:1977kg,Weinberg:1981eu,Weinberg:1979ma}. 
This will serve as reference point for the necessary generalizations for the computation
of the index for confined monopoles.  

First, one introduces the auxiliary ``Hamiltonian'' which factorizes the original operators,
\begin{equation}
  \label{H}
  H := \begin{bmatrix}  & -L^\dagger \\ L & \end{bmatrix} =:
  \Gamma^i\del_i + K(x)\, \quad\Rightarrow \quad 
   -H^2+M^2 =  \begin{bmatrix}   L^\dagger L+M^2& \\  & LL^\dagger+M^2\end{bmatrix} \, ,
\end{equation}
where we note that $(H+M)(-H+M) = -H^2+M^2$. The matrices $\Gamma^i$ are assumed to
satisfy a Clifford algebra, i.e. $\{\Gamma^i,\Gamma^j\}=\delta^{ij}$. Below we will 
discuss also situations where this is not case. The size and number of these
matrices is kept unspecified here. The relation on the right in
(\ref{H}) and the factorization given beneath, allows one to write the index function 
(\ref{ind1}), which is defined in terms of second order differential operators, 
as a functional of a first order operator. Restricting the trace in (\ref{ind1}) 
to the component and gauge indices one finds for the \emph{local} index function $\ca{J}(x,y,M)$:
\begin{align}
  \label{Iloc}
  G_M(x,y):=\frac{1}{H+M}&\ = (-H+M) \frac{1}{-H^2+M^2} \quad\Rightarrow\quad \nonumber\\[7pt]
  \ca{J}(x,y,M)&\ = \mathrm{Tr}\, \Big{\{}\, \Gamma_5\, \frac{M^2}{-H^2+M^2}\, \Big{\}} =   
                    \mathrm{Tr} \left\{\, \Gamma_5\, M\,  G_M(x,y)\, \right\}\, .
\end{align}
Here we have introduced the Green's function $G_M(x,y)$  and the chirality matrix 
$\Gamma_5 = \mathrm{diag}(\unit,-\unit)$. The symbol $\mathrm{Tr}$ includes now also 
the trace over the gamma matrices.
The global index function (\ref{ind1}) is then given by 
$\ca{I}(M)= \int_x  \ca{J}(x,y,M)|_{y=x}$, but it will be beneficial to keep the local expression.

Second, one uses the defining equations for the Green's function to rewrite the local index function 
$ \ca{J}(x,y,M)$ as a total derivative. 
The Green's function $G_M(x,y)$ satisfies the following first order equations, 
\begin{align}
  \label{gf}
  [\,H_x+M\,]\,G_M(x,y)&\ =\,\  [\,\Gamma^i\del_{x^i}+ K_x+M\,]\, G_M(x,y)\ \ =\, \delta(x-y) \unit 
   \nonumber\\[5pt]
  G_M(x,y)\, [\,H^\dagger_y+M\,]&\ =\, 
    G_M(x,y)\,[-\Gamma^i\overset{\leftarrow}{\del}_{y^i}+ K_y+M\,]\, =\, \delta(x-y) \unit\, ,
\end{align}
where we have indicated the coordinate dependence of the operators by an index and $\unit$ on the
r.h.s.\ stands for the identity in spinor- and gauge-matrix space. The off-diagonal
block structure of the operators imply that they anti-commute with the $\Gamma_5$ matrix, 
i.e.
\begin{equation}
  \label{g5}
  \{\Gamma_5,\, H\} = \{\Gamma_5,\, \Gamma^i\} = \{ \Gamma_5,\, K\} = 0\, .
\end{equation}
Therefore, taking the 
trace of the two equations in (\ref{gf}) with $\Gamma_5$  and adding the result gives,
\begin{align}
  \label{st}
   \ca{J}(x,y,M) =&\  -\fr{1}{2}\, (\del_{x^i} + \del_{y^i})\,\mathrm{Tr}
            \left\{\, \Gamma_5\, \Gamma_i\,  G_M(x,y)\, \right\} \nonumber\\[5pt]
            &\hspace{1.5cm} -\fr{1}{2}\, \mathrm{Tr} 
            \left\{\, \Gamma_5\, (K_x - K_y)\, G_M(x,y) \, \right\}\, ,       
\end{align}
where we have used the cyclicity of the trace in the finite-dimensional 
vector spaces, and that $\mathrm{Tr}\{\,\Gamma_5\,\unit\,\delta(x-y)\,\}=0$. This 
relation is of course problematic in the limit $y\rightarrow x$,
as is the second line in (\ref{st}) which seems to vanishing in this limit. These ambiguities 
are the source of possible anomalies and regularization is necessary for a proper treatment.
Putting this subtlety aside for a moment one sees that the index function 
$\ca{I}(M)= \int_x  \ca{J}(x,y,M)|_{y=x}$ is given by the integral of a total 
divergence and thus determined by a surface term, i.e. by the topological properties
of the fields in the operators $L$ and $L^\dagger$.

It has turned out that Pauli-Villars regularization is most 
convenient in the given context. This amounts to replacing the Green's function given 
in (\ref{Iloc}) by
\begin{equation}
  \label{Greg}
  G_M(x,y)\ \ \longrightarrow \ \ G^{\mathrm{reg}}(x,y) =  G_M(x,y) -  G_\mu(x,y)\, ,
\end{equation}
where the regulator mass $\mu$ is sent to infinity at the end. For the Pauli-Villars 
Green's function $G_\mu(x,y)$ the same relation as in (\ref{st}) holds and the difference of these
two equations yields
\begin{align}
  \label{jreg}
   \ca{I}(M)= \int_x \ca{J}(x,y,M)|_{y=x}& \nonumber\\ 
     &\hspace{-1cm}= \lim_{\mu\rightarrow\infty}\,\int_x\, \ca{J}(x,y,\mu)|_{y=x}
     -\fr{1}{2}\, \lim_{\mu\rightarrow\infty}\,\int_x\, \del_i\,j_i(x) \, ,
\end{align}
where the problematic term $\mathrm{Tr}\{\,\Gamma_5\,\unit\,\delta(x-y)\,\}$ has canceled. 
The regularized version of the last term in (\ref{st}), which is 
$\sim (K_x - K_y)\, G^{\mathrm{reg}}(x,y)$, rigorously vanishes for $y\rightarrow x$ since
$G^{\mathrm{reg}}(x,x)$ is well defined for every finite $\mu$. The 
potential anomaly has been shifted into the first term of (\ref{jreg}).
The current $j_i(x)$ is given by
\begin{equation}
  \label{j}
  j_i(x) = 
         \mathrm{Tr} \left\{\, \Gamma_5\, \Gamma_i\,  G^{\mathrm{reg}}(x,y)\, \right\}|_{y=x}\, .
\end{equation}
Finally we want to emphasize a point which has to be generalized in what follows. The
relation (\ref{st}), which eventually allows one to define the index function as a surface term, 
is obtained in this form thanks to the fact that the masses $M$ and $\mu$ are scalars that 
commute with all other quantities.\\

\noindent
{\bf{Anomaly vs.\ Anomaly.}} We briefly outline the well known relation between index theorems
and chiral anomalies. The purpose is to emphasize a particular observation concerning 
mass corrections and central charge anomalies for solitons (kinks), vortices and monopoles 
\cite{Goldhaber:2004kn, Rebhan:2006fg, Burke:2011mw}: As has been mentioned, 
the nontrivial mass corrections and
central charge anomalies stem from the spectral density (\ref{ind2}), which is non-vanishing only 
if the index function $\ca{I}(M)$ is \emph{not} independent of $M$. On the other hand, 
equation (\ref{jreg}) shows that the anomaly part, the first term on the r.h.s., is independent 
of $M$. We will argue now why the non/existence of a chiral anomaly in the auxiliary model associated
with the index under consideration, in general implies an anomaly/vanishing corrections for the 
solitonic objects in the actual model.
The Euclidean auxiliary model is:
  \begin{equation}
    \label{laux}
    S_E^{\mathrm{aux}} = \int\,d^Dx_E\, i\, \mathrm{Tr}\, \big{\{} \tilde{\Psi}^\dagger\, (H+M)\, \Psi \big{\}}
       \quad \Rightarrow\quad   G_M(x,y) = \langle \Psi(x)\, \tilde{\Psi}^\dagger(y)\rangle\, ,
  \end{equation}
where $G_M(x,y)$ is the Green's function defined in in (\ref{Iloc}), here given in terms
of the Euclidean two-point function of  the auxiliary system. The associated 
Minkowski action has a chiral symmetry, broken by the regulator mass $M$,  
whose current is\footnote{In Euclidean space $\Psi$ and $\tilde\Psi$ 
are independent, whereas going to Minkowski space implies $\tilde{\Psi}^\dagger\rightarrow \bar{\Psi}$,
see e.g. \cite{Smilga:2001ck}. The Euclidean and Minkowski space Green's function are related as 
$G_M(x^1,y^1,\ldots) = i\, G^{\mathrm{Mink}}_M(x^0,y^0,\ldots)|{{x^0 = -ix^1}\atop{y^0=-iy^1}}$. Hereby
is $G^{\mathrm{Mink}}_M = \langle \Psi\,\bar{\Psi}\rangle$ and $\bar{\Psi} = \Psi^\dagger i\Gamma^0$ 
is the Dirac conjugate spinor.} 
\begin{equation}
  \label{j5}
  j^\mu_{(5)} = i\, \mathrm{Tr}\, \big{\{}\bar{\Psi}\,  \Gamma^\mu\, \Gamma_5\,\Psi \big{\}}
               \quad \Rightarrow \quad
     \del_\mu\, j^\mu_{(5)} = 2i M\,  \mathrm{Tr}\, \big{\{} \bar{\Psi} \Gamma_5 \Psi \big{\}}
                            - \mathrm{a}_{(5)}(x)\, .
\end{equation}
The second relation is the anomalous conservation equation,
including the explicit breaking term, for the chiral current. 
Taking the expectation
value of the anomalous conservation equation 
one sees, after Wick rotation $x^0=-ix^1, \Gamma^0=-i\Gamma^1$, that the first term on the r.h.s. 
is equal to $- 2\, \mathrm{Tr} \left\{\, \Gamma_5\, M\,  G_M\, \right\} = -2\, \ca{J}(x,x,M)$, 
see (\ref{Iloc}). Hence the index function (\ref{jreg}) can alternatively be written as
\begin{align}
  \label{an}
 \ca{I}(M)= \int_x \ca{J}(x,x,M) = - \fr{1}{2} \int_x \mathrm{a}_{(5)}(x)
     -\fr{1}{2}\,\int_x\, \del_i\,\langle\, j^i_{(5)}\rangle \, . 
\end{align}
Given the definition of the chiral current (\ref{j5}) and the propagator in (\ref{laux}) one sees
that the expectation value of the chiral current is identical with the above 
encountered current (\ref{j}), i.e.\ after Wick rotation is $\langle\, j^i_{(5)}\rangle = j^i(x)$. 
As it is well known, 
the chiral anomaly is associated with the index of the Euclidean Dirac operator $\slashed{D}_E:= H$ of the 
Lagrangian (\ref{laux}). Explicitly one has
\begin{equation}
  \label{indaux}
  \mathrm{index} (\slashed{D}_E) = \ca{I}(0) = -\frac{1}{2}\,\int\, \mathrm{a}_{(5)}(x)\, ,
\end{equation}
see for example \cite{Bilal:2008qx} for a recent detailed account.  This means that if the auxiliary 
system (\ref{laux}) has a chiral anomaly the index function $\ca{I}(M)$ is completely determined by it
and there are no contributions from the surface term\footnote{The 
only way around this would be to assume that the surface term contribution vanishes only in the limit 
$M\rightarrow 0$. This however leads to a contradiction with (\ref{ind2}).} 
in (\ref{jreg}). As mentioned above, the anomaly
term is independent of the IR-regulator mass $M$ and consequently the continuum contribution 
(\ref{ind2}) and thus the spectral density $\Delta\rho$ vanishes in this case. 

The quantum mass corrections and central charge anomalies for solitons are determined
by the spectral density $\Delta\rho$. Hence, the arguments given here confirm the observed fact
\cite{Goldhaber:2004kn, Rebhan:2006fg, Burke:2011mw},
that if the auxiliary system in the soliton background is anomalous the quantum corrections for
the soliton mass and central charge anomaly vanish, and vice versa. 

In particular, since there are no chiral anomalies in odd dimensions,  one finds in general
nontrivial mass corrections and anomalous central charges for solitons that occupy 
one and three spatial dimensions like kinks and monopoles. An exception to the rule occurs 
when the field content leads to a vanishing overall factor, as it is the case for conformal models like
$\sN =4$ supersymmetric YM-theory \cite{Rebhan:2006fg} or $\sN =2$ SQCD with $2N = N_f$ 
\cite{Burke:2011mw}.

\subsection{Overall anomaly}\label{secanotot}

Since the considered setting is three-dimensional the anomaly part in (\ref{jreg}) vanishes. 
We sketch here the proof of this statement and give a more detailed account in appendix \ref{anotot}.
The anomaly 
\begin{equation}
  \label{overan}
  \mathcal{A}^{\mathrm{tot}} = \int_x\,\ca{J}^{\mathrm{ano}}(x,x) = \lim_{\mu\rightarrow\infty}\,\int_x\, \mathrm{Tr}\,
                       \big{\{} \Gamma_5 \frac{\mu^2}{-H^2+\mu^2}\big{\}}|_{y=x}\, ,
\end{equation}
is easily evaluated by expanding the Green's function $(-H^2+\mu^2)^{-1}$  for large $\mu$.
The notation $\mathcal{A}^{\mathrm{tot}}$ indicates that this is the total anomaly 
of the system, below we will find a different and essential sub-anomaly.
For the given fluctuation operators (\ref{LLd}) $H$ is of the form (\ref{H}) with
\begin{equation}
  \label{K}
  \Gamma^i = \begin{bmatrix} & \unit_2\otimes \sigma^i \\ \unit_2\otimes\sigma^i &
             \end{bmatrix} \ ,\ \
  K(x) = -i\,\begin{bmatrix}  & \, Q \,\\
                         \, Q^\dagger & 
              \end{bmatrix} \ ,\ \
     Q = \begin{pmatrix}  & &\bar{\slashed{\ca{A}}}^{\mathrm{a}} & -g \bar\Sigma_i^{\mathrm{r}}\, \\
                              & & g \Sigma_i^{\mathrm{r}} & {\slashed{\ca{A}}}^{\mathrm{f}}
          \end{pmatrix} \, .
\end{equation}
The expansion for large $\mu$ reads as
\begin{equation}
  \label{anex}
   \frac{1}{-H^2+\mu^2} = \Delta_\mu(x-y) + \sum_{n\geq 1} \int_{z_1,\ldots z_n} \hspace{-8mm}\Delta_\mu(x-z_1) 
                        \ca{O}(z_1)\Delta_\mu(z_1-z_2) \ldots \ca{O}(z_n)\Delta_\mu(z_n-y),  
\end{equation}
with $\Delta_\mu$ being the three-dimensional propagator (\ref{massprop}). 
The first term obviously vanishes under trace with $\Gamma_5$ and the insertion $\ca{O}$ is the deviation 
from the free Laplacian,
\begin{equation}
  \label{O}
 \ca{O} = H^2 - \del^2 = \Gamma^i\del_iK + K^2 + 2 K_i\del_i\, .
\end{equation}
Here $K_i=\{\Gamma_i,K\}$, which is of  form $\mathrm{diag}(a,a)$ such that 
$\mathrm{Tr}\,\{ \Gamma_5 K_{i_1}\ldots K_{i_n}\}=0$ for any $n$. In order to estimate which orders in the 
expansion (\ref{anex}) can contribute in the limit ${\mu\rightarrow\infty}$ 
we note that the $n$'th term scales as  $\mu^{(1-2n+r)}$ for $y=x$, with $r$ being the number 
of derivatives of $\ca{O}$ that appear in the product\footnote{Changing the integration 
variables as $z_i \rightarrow x+z_i/\mu$ one sees that the 
coefficient functions of $\ca{O}(z_i)$ can be evaluated at $x$. An expansion introduces negative powers of $\mu$
while the $z$-integrations are finite due to the exponential decay of the propagator.}.  
The term with the maximum number of derivatives,
$r=n$, vanishes after taking the trace because of the just mentioned property of $K_i$. Hence,
the terms that can contribute are $(n=1, r=0)$ and $(n=2, r=1)$. Both terms are regular for $y=x$ and 
therefore the second term with the single derivative vanishes by symmetric integration.
Thus the only a priori surviving (and potentially divergent) term is
\begin{equation}
  \label{jan}
  \ca{J}^{\mathrm{ano}}(x,x) =\lim_{\mu\rightarrow \infty}\, 
        \left[\,\frac{\mu}{8\pi}\, \mathrm{Tr}\,\{\,\Gamma_5 \Gamma^i\del_iK(x)\,\}\right]\, ,
\end{equation}
which vanishes due to the fact that $\mathrm{Tr}\,\{\,\Gamma_5 \Gamma^i K(x)\,\} = 
2\, \mathrm{tr}\ (\,\ca{A}_4^{\mathrm{a}}-\ca{A}_4^{\mathrm{f}}\,)\,\mathrm{tr}\,\sigma^i = 0$.
 
Thus the overall anomaly contribution 
to the index function $\ca{I}(M)$ vanishes, $\mathcal{A}^{\mathrm{tot}} = 0$, as expected from the 
discussion following (\ref{laux}). Consequently, if non-vanishing, $\ca{I}(M)$ will be $M$-dependent and
lead to a non-vanishing spectral density. 

\subsection{Index of a boundary term}

Having shown that the anomalous contribution vanishes the index function 
(\ref{jreg}) reduces to the surface term:
\begin{align}
  \label{jsurf}
   \ca{I}(M)=&\  -\fr{1}{2}\, \lim_{\mu\rightarrow\infty}\,\int_x\, \del_i\,j_i(x)\nonumber\\[3pt]
        &\ -\fr{1}{2}\, \lim_{\mu\rightarrow\infty}\,\Big{[} \int
            d\theta\, dx^3 \ r\, j_r\,|_{r=\infty} \ \ 
            + \int d\theta\, rdr \ j_3\, 
            \Big{|}_{x^3=-\infty}^{x^3=\infty}\ \Big{]}\nonumber\\[3pt]
    &=:\ \ca{I}_{\mathfrak{Z}_\infty}(M) + \big{[}\,\ca{I}_{\mathfrak{D}_+}(M) 
                    - \ca{I}_{\mathfrak{D}_-}(M)\,\big{]}\, ,
\end{align}
where $r, \theta, x^3$ are the cylindrical coordinates. 
In the last line we introduced a notation in accordance with the definitions of the boundary at 
infinity in figure \ref{fig}.  To evaluate the last two terms in  
(\ref{jsurf}) one has to  know the field configuration on the whole discs 
at infinity $\mathfrak{D}_\pm$, which describe the full vortex dependence and
is not even known analytically. We will now show  how to resolve this situation and 
how to reduce the computation of the index to the topological properties of the field configuration,
as it is the case for more familiar situations.

\subsubsection*{Cylinder Wall $\mathfrak{Z_\infty}$:} 
For the given situation this contribution is easy to 
evaluate. With $(x^\alpha) = r\, (\cos\theta,\sin\theta)$ one has,
\begin{equation}
  \label{Iz}
  \ca{I}_{\mathfrak{Z}_\infty}(M) = -\fr{1}{2}\, \lim_{\mu\rightarrow\infty}
        \int d\theta\, dx^3\, x^\alpha j_\alpha\,|_{r=\infty} \quad \textrm{with} \quad 
    j_\alpha = \mathrm{Tr}\, \big{\{}\, \Gamma_5\Gamma_\alpha [\frac{1}{H + M} - (\mu)\,]\,\big{\}},
\end{equation}
where we indicated the subtraction of the same term with $M$ replaced by $\mu$. To compute this
quantity we first note that with the asymptotic behavior as discussed around the equations
(\ref{asym1}) and (\ref{asym2}) the building blocks (\ref{ddbps}) simplify considerably. 
Hence the operators (\ref{LL}) take the simple ultra-diagonal form,
\begin{equation}
  \label{LLz}
  L^\dagger L \approx L L^\dagger \approx \begin{bmatrix}
       \unit_2\otimes [\, -\ca{D}_{\ww k}^{\mathrm{a}\,2} + g^2 (\Sigma_i\otimes\bar\Sigma_i)^{\mathrm{r}}\,] &\\
        &  \unit_2\otimes [\, -\delta_{ij} \ca{D}_{\ww k}^{\mathrm{f}\, 2} 
                    + g^2 (\bar\Sigma_j\Sigma_i)\,]\end{bmatrix}\, ,
\end{equation}
where $\approx$ again stands for equal up to exponentially suppressed terms $\sim e^{-gvr}$.  The relevant 
term for computing $j_\alpha\,|_{r=\infty}$ therefore becomes,
\begin{align}
  \label{jr}
    \mathrm{Tr}\, \big{\{}\, \Gamma_5\Gamma_\alpha \frac{1}{H + M}\,\big{\}}\,|_{r=\infty}
      \ =   \mathrm{Tr}\, \big{\{}\, \Gamma_5\Gamma_\alpha (-H)\frac{1}{-H^2 + M^2}\,\big{\}}\,|_{r=\infty}
      \ \sim \mathrm{tr}\, \sigma^\alpha = 0 \, .
       % \ca{A}^{\mathrm{a}}_4 \,\frac{1}{-\ca{D}_{\ww k}^{\mathrm{a}\,2} 
        %                     + g^2 (\Sigma_i\otimes\bar\Sigma_i)^{\mathrm{r}}} 
      %- \delta_{ij}\, \ca{A}^{\mathrm{f}}_4 \,\frac{1}{ -\delta_{ij} \ca{D}_{\ww k}^{\mathrm{f}\, 2} 
       %             + g^2 (\bar\Sigma_j\Sigma_i)}\, \}\, 
\end{align}
Consequently there is no contribution from the cylinder wall, i.e. 
$\ca{I}_{\mathfrak{Z}_\infty}(M) = 0$.

\subsubsection*{The Disc contribution at $\mathfrak{D}_{\pm}$:}

With the anomaly and the contribution at the cylinder wall $\mathfrak{Z_\infty}$
vanishing the index function  (\ref{jsurf}) is entirely given by the 
contribution  from the infinite discs 
$\mathfrak{D}_\pm$ at $x^3\rightarrow \pm\infty$:
\begin{equation}
  \label{discIM}
  \ca{I}(M)= \ca{I}_{\mathfrak{D}_+}(M)  - \ca{I}_{\mathfrak{D}_-}(M)\, ,
\end{equation}
where the explicit form of the disc contributions is with  (\ref{jsurf}), (\ref{j}): 
\begin{align}
  \label{jD}
   \ca{I}_{\mathfrak{D}_\pm} (M)&\ = -\fr{1}{2}\, \lim_{\mu\rightarrow\infty}\,
             \int_{\mathfrak{D}_\pm} \ j_3,\nonumber\\[5pt]
    j_3&\  = \mathrm{Tr}\, \big{\{}\, \Gamma_5\Gamma_3 [\frac{1}{H + M} - (\mu)\,]\,\big{\}}|_{y=x}
     =: j^{\mathrm{reg}}(x,y)|_{y=x}\, .
\end{align}

With the asymptotic behavior at the discs $\mathfrak{D}_\pm$ 
as described around (\ref{asym3}) and the properties for the fundamental building blocks (\ref{ddbps})
the operators $L^\dagger L$ and $L L^\dagger$ have the following block structure,
\begin{equation}
  \label{block1}
   L^\dagger L \approx L L^\dagger \approx \left[
    \begin{smallmatrix} \ast & & & \\  & \ast & \ast & \\ & \ast & \ast & \\ & &  & \ast \end{smallmatrix}
     \right]\, . 
\end{equation}
Here and in the following $\approx$ stands for equal up to terms that are exponentially suppressed
as $e^{-|\Delta x \Delta m|}$, see the discussion around (\ref{asym3}). We therefore perform a 
transformation such that,
\begin{align}
  \label{otrans}
   -H^2 = \begin{bmatrix}L^\dagger L& \\ & L L^\dagger\end{bmatrix} \longrightarrow 
          - (O\,H\,O^t)^2 = \left[\begin{smallmatrix} \ast & & & & & & & \\ &\ast &  & & & & & \\
                            & &\ast &  & & & & \\  & & & \ast &  & & & & \\ 
                             & & & & \ast & \ast & & \\  & & & & \ast & \ast & &\\
                              & & & & & & \ast & \ast\\  & & & & & & \ast & \ast
                           \end{smallmatrix}\right]\, ,
\end{align}
where $O$ is orthogonal\footnote{\label{fto}The explicit form of the transformation is 
$O = \Big{[} \begin{array}{cc} a & \\[-7pt] 
                       & a\\[-4pt] 
                     b &  \\[-7pt]
                       & b
        \end{array}\Big{]}$ with $a = (\begin{smallmatrix}1&0&0&0\\0&0&0&1\end{smallmatrix})$ and
$b = (\begin{smallmatrix}0&1&0&0\\0&0&1&0\end{smallmatrix})$ .}. The resulting
harmonized block structure is described in terms of the following substructure of operators:
\begin{align}
  \label{lld}
\ca{L} = \begin{bmatrix} - \ca{D}_z^{\mathrm{a}} &\, {\scriptstyle{+}} i g \bar\Sigma_i^{\mathrm{r}}\, \\
                                \,{\scriptstyle{+}} i g \Sigma_i^{\mathrm{r}} & \ca{D}_{\bar{z}}^{\mathrm{f}}
                \end{bmatrix} \ , \qquad
\ca{L}^\dagger = \begin{bmatrix} \ca{D}_{\bar{z}}^{\mathrm{a}} & 
                                         \, {\scriptstyle{-}} i g \bar\Sigma_i^{\mathrm{r}}\,\\
                                \,{\scriptstyle{-}} i g \Sigma_i^{\mathrm{r}} & -\ca{D}_{z}^{\mathrm{f}}
                \end{bmatrix}\, , 
\end{align}
where from now on it is always understood that all objects are considered to be at the 
discs $\mathfrak{D}_{\pm}$. Another structure which will be important in the following is given by the operator
\begin{equation}
  \label{massop}
  \mop = \begin{pmatrix} \sqrt{2}\ \phi_0^{\mathrm{a}} & \\ & \sqrt{2}\, (\phi_0 + m_i) \end{pmatrix}\, , \qquad
\end{equation}
where $\phi_0 =  \phi_0{}^A T^A$ is the vacuum  (\ref{vac1}) 
which is a matrix in the fundamental representation
and $(\phi_0 + m_i)$ is the operator when acting on a field $y_i$ in the fundamental representation 
which carries a flavor index $i$.
The operator  $\phi_0^{\mathrm{a}}$ is the associated adjoint representation, i.e. it acts as 
commutator or written as a matrix  it is of the form  $(\phi_0^{\mathrm{a}})^A{}_B = -i\, \phi_0{}^C f^{CAB}$.

The operators introduced in (\ref{lld}) and (\ref{massop}) satisfy the following 
relations:
\begin{align}
  \label{lmrel}
  [\,\ca{L}\, ,\, \mop\,]  \approx  [\,\ca{L}\, ,\, \del_3\,] \approx 0 \, ,
\end{align}
with $\ca{L}^\dagger$ satisfying the same relations and we note that $ \ca{L}^\dagger\ca{L}$ is diagonal. 
These relations follow from the asymptotic 
BPS equations given below (\ref{asym3}) and the fact that the $x^3$ dependence is exponentially 
located at the position of the monopoles, which are assumed to be infinitely far away from the 
boundary. 

Transforming also the gamma matrices in (\ref{jD}) with the orthogonal transformation $O$
the current on the discs $\mathfrak{D}_\pm$ takes the form,
\begin{align}
  \label{dcur}
 j(M,x,y)& =\ \mathrm{Tr}\, 
     \big{\{} \Gamma_5'\Gamma_3' \, \frac{1}{H' + M} \,\big{\}}
    =   \mathrm{Tr}\, \big{\{} \Gamma_5'\Gamma_3'\, (-H'+ M)\, \frac{1}{-H'^2 + M^2}\,\big{\}}
    \nonumber\\[7pt]  
   &=\ 
    2\, \mathrm{Tr}\, \big{\{}\, \mop\, \Big{[}\,\frac{1}{\ca{L}^\dagger\ca{L} -\del_3^2 + \mop^2 + M^2} 
          - \frac{1}{\ca{L}\ca{L}^\dagger -\del_3^2 + \mop^2 + M^2} \,\Big{]}\, \big{\}}.
\end{align}
The actual quantity needed to compute $\ca{I}_{\mathfrak{D}_\pm} (M)$ is 
$j^{\mathrm{reg}}(x,y) = j_3(M,x,y) - j_3(\mu,x,y)$, see (\ref{jD}). 

The quantity (\ref{dcur}) reminds also of our starting point, the definition  of the index function 
(\ref{ind1}). It is this similarity that we are going to use to compute (\ref{dcur}). The difference
to (\ref{ind1}) is that in addition to the parameter $M$ it contains the operator $\mop$ and especially the 
derivative operator $\del_3$, which contrary to the operator $\cal{L}$, does not act 
in the bulk of the discs $\mathfrak{D}_\pm$, but orthogonal to them. Besides 
these subtleties, which we address in a moment, the quantities  $j(M,x,y)$ may be considered 
as the definition of a new index function, analogous to (\ref{ind1}) or more accurately analogous to the 
local index function $\ca{J}(x,y,M)$ in (\ref{Iloc}), and we will compute 
it in this spirit.

We first recall the basic properties that were used in section \ref{princip} to formulate
the index function of the form (\ref{ind1}) as the sum of an anomaly and surface term:
i.) The factorization property which allows to express the index function in terms of
a Green's function of a first order operator (\ref{Iloc}) which satisfies equations of the form 
(\ref{gf}), and ii.) commuting properties with a chiral matrix $\Gamma_5$ (\ref{g5}) which lead
to (\ref{st}). In this last step it is also important that $M$ commutes with all other 
expressions.

In a first step one has to deal with the presence of the unusual derivative operator $\del_3$ 
in (\ref{dcur}) which acts orthogonal to the boundary and prevents a proper factorization 
mentioned under i). To this end we introduce the Fourier transformed 
current $\tilde{j}$ in the following
way:
\begin{align}
  \label{jred}
   j(M,x,y) =&\   \int\, \frac{dp}{2\pi}\ e^{ip(x^3 - y^3)}\ \tilde{j}(M_p,x^\alpha,y^\alpha) 
    \qquad \textrm{with}\nonumber\\[5pt]
   \tilde{j}(M_p) :=&\ 
   2\, \mathrm{Tr}\, \big{\{}\, \mop\, \Big{[}\,\frac{1}{\ca{L}^\dagger\ca{L}  + \mop^2 + M^2_p} 
          - \frac{1}{\ca{L}\ca{L}^\dagger+ \mop^2 + M^2_p} \,\Big{]}\, \big{\}}\, ,
\end{align}
where in the first line we indicated explicitly the coordinate dependence with $\alpha =1,2$
denoting the coordinates on the discs $\mathfrak{D}_\pm$. From now on 
we will suppress this notation, as in the second line of (\ref{jred}). The momentum dependent IR regulator
is $M^2_p = M^2+p^2$. Using a complete eigen system of the (diagonal) operator $\cal{L}^\dagger\cal{L}$
and the fact that it is isospectral to the operator $\cal{L}\cal{L}^\dagger$ (zero-modes are treated
without problems separately) it is easy to prove the identity (\ref{jred}).

We follow now the steps of section \ref{princip} which allowed one to express the 
original local index function $\ca{J}(x,y,M)$ as a surface term and an anomaly.
First one has to find the proper factorization in terms of an auxiliary Hamiltonian. 
Defining the operators 
\begin{align}
  \label{HH}
   \ca{H} =&\ \begin{bmatrix}
        \unit_2\otimes \mop+  \sigma^3\otimes i M_p \unit_2 &  -\unit_2\otimes \ca{L}^\dagger\\
          \unit_2\otimes\cal{L} &    \unit_2\otimes \mop - \sigma^3\otimes  i M_p\unit_2
           \end{bmatrix}\nonumber\\[5pt]
   \ca{\tilde H} =&\ \begin{bmatrix}
        -\unit_2\otimes \mop+ \sigma^3\otimes i M_p\unit_2 &  -\unit_2\otimes \ca{L}^\dagger\\
          \unit_2\otimes\cal{L} &   - \unit_2\otimes \mop - \sigma^3\otimes i M_p\unit_2 
           \end{bmatrix}\, ,
\end{align}
one can rewrite the Fourier transformed current (\ref{jred}) in the desired form:
\begin{align}
  \label{jred2}
    \tilde{j}(M_p) = 
    \mathrm{Tr}\, \big{\{}\, \Gamma_5\, \ca{H}\, \frac{1}{-\tilde{\ca{H}}\, \ca{H}}\, \big{\}} =   
                   - \mathrm{Tr}\, \big{\{}\, \Gamma_5\, \tilde{G}_{M_p} \big{\}}\qquad
      \textrm{with:} \quad  \tilde{G}_{M_p} := \frac{1}{\tilde{\ca{H}}}\, .
\end{align}
This is analogous to (\ref{Iloc}). We now have to find the
generalization of the steps which led from (\ref{gf}) to (\ref{st}) for 
$\tilde{j}(M_p)$ and $\tilde{j}(\mu_p)$ such that we eventually obtain a 
regularized expression analogous to (\ref{jreg}). The auxiliary Hamiltonian is of the form
\begin{equation}
  \label{Htilde}
  \tilde{\ca{H}} = \Gamma^\alpha\del_{\alpha}  + K(x) + K_{M_p} \quad \textrm{with}\quad  
     K_{M_p} = -\unit_4\otimes \mop + \sigma^3\otimes\sigma^3\otimes i M_p\unit_2\, ,
\end{equation}
whereas\footnote{These matrices, though of the same block structure,  
differ from the ones of section \ref{secanotot}. They are defined via (\ref{HH}).} 
$\Gamma^\alpha$, $K(x)$ are independent of $\mop$ and $M_p$ and are block off-diagonal
so that they anti-commute with $\Gamma_5$. The operator $\ca{H}$ is of the same form 
with $K_{M_P}\rightarrow - K_{M_P}^\dagger$.

Second, we again use the defining equations for the Green's function to rewrite
now $\tilde{j}(M_p)$ as a surface term. The 
Green's function (\ref{jred2}) satisfy the equations 
\begin{align}
  \label{gftild}
   [\,\Gamma^\alpha\del_{x^\alpha}+ K_x+ K_{M_p}\,]\, \tilde{G}_{M_p}(x,y) =&\ \delta(x-y) \unit\, , 
   \nonumber\\[5pt]
\tilde{G}_{M_p}(x,y)\,[-\Gamma^\alpha\overset{\leftarrow}{\del}_{y^\alpha}+ K_y+ K_{M_p}\,] =&\ \delta(x-y) \unit\, ,
\end{align}
which is similar to (\ref{gf}).
However, the simple IR regulator $M$ has been replaced by operator $K_{M_p}$. Above 
we took the trace with $\Gamma_5$ of  (\ref{gf}) and the sum led to (\ref{st}).
Due to the presence of the operator $K_{M_p}$ one has to generalize this procedure as follows. We assume 
the existence of a block diagonal matrix $\Gamma$ (given explicitly below) such that it commutes with $K_{M_p}$ and has similar 
commutation properties as $\Gamma_5$ before (\ref{g5}), i.e.\footnote{Actually, this relations are stronger than
really needed. It suffice that they hold when inserted in the trace of $\tilde{G}_{M_p}$}
\begin{equation}
  \label{gstar}
  \{\Gamma\,,\, \Gamma^\alpha\} = \{\Gamma\,,\, K\} = 0  =  [\,\Gamma\,,\, K_{M_p} ] \, .  
\end{equation}
The vanishing of the last commutator is the analog of $[\, \Gamma_5\, ,\, M]=0$ in the previous derivation.
Taking the trace with $\Gamma$  of the sum of the two equations in  (\ref{gftild})
one obtains
\begin{align}
  \label{sttild}
    \mathrm{Tr}\,\{\Gamma K_{M_p} \tilde{G}_{M_p}\} =&\   
    -\fr{1}{2} (\del_{x^\alpha} + \del_{y^\alpha})\,\mathrm{Tr}\,\{\, \Gamma\Gamma_\alpha\, \tilde{G}_{M_p}\}
    \nonumber\\[3pt]
  &\hspace{1cm} -\fr{1}{2} \mathrm{Tr}\, \{\,\Gamma (K_x-K_y)\, \tilde{G}_{M_p} \} + 
         \delta(x-y)\,\mathrm{Tr}\,\Gamma \, . 
\end{align}
The l.h.s.\ should now give $\tilde{j}(M_p)$ from (\ref{jred}). To this end we set  
\begin{equation}
  \label{gstar2}
  \Gamma = \begin{bmatrix}\unit_2 & \\ 
           & -\unit_2 \end{bmatrix}\otimes\mop\,\frac{1}{\mop^2+ M_p^2}
   \quad \ \  \Rightarrow \quad \ \ 
    \mathrm{Tr}\,\{\Gamma K_{M_p} \tilde{G}_{M_p}\} = \tilde{j}(M_p)\, ,
\end{equation}
which can be easily seen using (\ref{jred2}).
Because of relations (\ref{lmrel}) this $\Gamma$  satisfies the 
assumed commutation relations (\ref{gstar}). An important property of the nontrivial 
factor in $\Gamma$ is that it exists even in the limit $M_p \rightarrow 0$, since the zero-eigen values of
$\ca{M}^2$ are projected out by the nominator  $\ca{M}$.

The quantity which is actually needed  is not $\tilde{j}(M_p)$ but the regularized 
expression $\tilde{j}^{\mathrm{reg}}= \tilde{j}(M_p) - \tilde{j}(\mu_p)$, see
(\ref{jD}). One finds the same relation (\ref{sttild}) for $\tilde{G}_{\mu_p}$ by replacing
$M_p$ by $\mu_p$ in the above derivation. The last relation in (\ref{gstar2}) does however 
not hold in this case due to the form of $\Gamma$. Subtracting from (\ref{sttild}) the 
analogous equation for  $\tilde{G}_{\mu_p}$ one finds
\begin{align}
  \label{streg}
 \mathrm{Tr}\, \{\,\Gamma\, (K_{M_p} \tilde{G}_{M_p} - K_{\mu_p} \tilde{G}_{\mu_p})\}=
   &\ -\fr{1}{2}(\del_{x^\alpha} + \del_{y^\alpha})  
   \mathrm{Tr}\, \{\,\Gamma\,\Gamma_\alpha(\tilde{G}_{M_p} - \tilde{G}_{\mu_p})\} \nonumber\\[3pt]
  &\quad\quad\quad-\fr{1}{2}\mathrm{Tr}\, \{\,\Gamma\,(K_x-K_y)(\tilde{G}_{M_p} - \tilde{G}_{\mu_p})\}\,.
\end{align}
The terms on the r.h.s.\ are regular for $y\rightarrow x$ since only the regularized Green's function 
$\tilde{G}^{\mathrm{reg}}=\tilde{G}_{M_p} - \tilde{G}_{\mu_p}$ appears. For this to be the case it is important 
to use the same matrix $\Gamma$ for both cases of the relation (\ref{sttild}), either with parameter $M_p$ or 
$\mu_p$. The second term on the r.h.s.\  of (\ref{streg}) vanishes therefore in the limit  $y\rightarrow x$ 
and will be not considered further.

The l.h.s.\ of (\ref{streg}) is not yet of the desired form to give $\tilde{j}^{\mathrm{reg}}$ and
also it seems that the usual anomaly term is missing. Using the explicit form (\ref{Htilde}) of 
$K_{\lambda=M_{p},\,\mu_p}$ and (\ref{jred2})  the l.h.s.\ of (\ref{streg}) can be written as:
\begin{equation}
  \label{lhs}
    \mathrm{Tr}\, \{\,\Gamma\, (K_{M_p} \tilde{G}_{M_p} - K_{\mu_p} \tilde{G}_{\mu_p})\}
    \,=\, \tilde{j}^{\mathrm{reg}}\, -  \,
      \mathrm{Tr}\, \{\, \unit_4\otimes \frac{M^2-\mu^2}{\ca{M}^2+M_p^2}\, \Gamma_5\, \tilde{G}_{\mu_p}\}\, .  
\end{equation}

Consequently, one can rewrite the surface term contribution to the index function $\ca{I}(M)$,
which comes exclusively from the discs $\mathfrak{D}_\pm$, as a surface term at the boundary 
of the discs $\del\mathfrak{D}_\pm$, where all fields assume values in the vacuum moduli space 
plus an anomaly term:
\begin{align}
  \label{sst}
     \ca{I}_{\mathfrak{D}_\pm} (M) &\ = -\fr{1}{2}\, \lim_{\mu\rightarrow\infty}\,
             \int_{\mathfrak{D}_\pm}  \int\, \frac{dp}{2\pi} \, 
              \big{[}\, \del_\alpha\, q_\alpha (x) + \mathrm{a}(x) \,\big{]}\quad\quad \textrm{with:} 
        \nonumber\\[10pt]
      q_\alpha (x) &\ =\, -\fr{1}{2} 
           \mathrm{Tr}\, \{\,\Gamma\,\Gamma_\alpha\,\tilde{G}^{\mathrm{reg}}\}|_{y=x}\ \ , \ \ \ \
      \mathrm{a}(x)  \,=\, 
 \mathrm{Tr}\, \{\, \unit_4\otimes \frac{M^2-\mu^2}{\ca{M}^2+M_p^2}\, \Gamma_5\, \tilde{G}_{\mu_p}\}|_{y=x}\ . 
\end{align}
We state here only the results for the evaluation of the expressions in (\ref{sst}). 
The proofs for the following statements are given in appendix \ref{das}. 
Both objects, $q_\alpha(x)$ and $\mathrm{a}(x)$, have a prefactor $\ca{M}(\ca{M}^2 + M^2_p)^{-1}$
inside the trace\footnote{For $q_\alpha(x)$ it is explicitly contained in $\Gamma$ (\ref{gstar2}).
For $\mathrm{a}(x)$ it follows from the first equation in (\ref{jred}) in combination with the already present 
prefactor, see (\ref{a}).} which is of utmost importance and represents the modification of the expressions
 (\ref{Iloc}) and (\ref{j}) that enter in standard formula for the index function $\ca{I}(M)$ 
(\ref{jreg}). This factor renders the $p$-integration finite and projects out any poles and massless fields
in the propagators that might appear in the $M\rightarrow 0$ limit, see the comments below (\ref{gstar2}) and
appendix \ref{das}.

It turns out that only the anomaly contributes in (\ref{sst}) and the result is
\begin{align}
  \label{indres}
   \ca{I}(M)\
            =\ \ca{A}^{\mathrm{disc}}\ =\ -\frac{1}{4\pi} \int_{\mathfrak{D}_{+} - \mathfrak{D}_{-}}\hspace{-4mm}
                          \mathrm{Tr}\, \Big{\{} \frac{\ca{M}}{\sqrt{\ca{M}^2+M^2}}
                 \begin{bmatrix} \ca{B}_3^{\mathrm{a}}&\\[-4pt] 
                           & - \delta_{ij} \ca{B}_3^{\mathrm{f}}\end{bmatrix} \Big{\}}  \, ,
\end{align}
where we used that $\ca{I}= \ca{I}_{\mathfrak{D}_+}  - \ca{I}_{\mathfrak{D}_-}$, see (\ref{discIM}). 
Remarkably, and for the non-vanishing of the spectral density (\ref{ind2}) essential, 
this anomaly is \emph{not} independent of the IR-regulator $M$.

\subsection{Topological charges}

We describe now the relation to topological quantities for the index (\ref{indres}). 

\subsubsection*{Topological charges:}

In order to define define certain topological invariants we consider the classical
energy of the  field configurations  considered here. The BPS equations (\ref{BPS}) imply 
that the (classical) energy of confined monopoles is given in terms of the total vortex number 
and the magnetic charge of the monopole \cite{Burke:2011mw}:
\begin{equation}
  \label{Ecl}
  E_{\mathrm{cl}} = 2\pi v^2 \, k^{\mathrm{vor}} L 
    \ + \ \fr{2}{g^2}\int d^3x\ \del_k\,\mathrm{Tr}\, \{\sqrt{2}\ \varphi\, \ca{B}_k\, \}\, ,
\end{equation}
where the first term is the total vortex tension times the regulated extent of size $L$ in the $x^3$ direction. 
The total vortex number is $k^{\mathrm{vor}} = \frac{1}{2\pi}\int d^2x \mathrm{Tr}\,\ca{B}_3 =\mathrm{Tr}\, w$, 
see (\ref{asym2}). However, in the following the second term, which gives the energy of the monopole and
encodes the magnetic charges will be of prime interest. 

For the given asymptotic behavior (\ref{asym2}) the surface term in (\ref{Ecl}) reduces to contributions 
from magnetic flux through the discs $\mathfrak{D}_\pm$. Following \cite{Nitta:2010nd} one
can express this flux in terms of the individual contributions to the vortex number according to 
the symmetry breaking pattern  (\ref{vac4}). Corresponding to the group $H_{C+F}$ (\ref{HCF}) there 
are $q$ distinct topological  quantum numbers
\begin{equation}
  \label{ktop}
  \Delta k_r^{\mathrm{vor}} = \frac{1}{2\pi}\int_{\mathfrak{D}_{+} - \mathfrak{D}_{-}} \hspace{-3mm} 
     \mathrm{Tr}\,\{\ca{B}_3\, t_r^0\,\}
    \quad\Rightarrow\quad \sum_{r=1}^q  \Delta k_r^{\mathrm{vor}} =  k^{\mathrm{vor}}- k^{\mathrm{vor}} =0 \ ,
\end{equation}
where $t_r^0= \mathrm{diag}(0,\ldots, \unit_{n_r},\ldots0)$ is the $U(1)$ generator of the $r$'th 
factor of  $H_{C+F}$ (\ref{HCF}). The vacuum value of the adjoint scalar 
(\ref{vac1}) can thus be written as $\phi_0 = - \sum_{r=1}^q m_r\, t_r^0$.  
The monopole mass contribution to the classical energy 
is then given by
\begin{equation}
  \label{Mcl}
  M^{\mathrm{mon}}_{cl} \ =\   
    \fr{2}{g^2}\int_{\mathfrak{D}_{+} - \mathfrak{D}_{-}} \hspace{-3mm}
  \mathrm{Tr}\, \{\sqrt{2}\ \phi_0\, \ca{B}_3\, \} \ = \
     -\frac{4\pi}{g^2} \sum_{r=1}^q \sqrt{2}\, m_r \Delta\,k_r^{\mathrm{vor}}\,,
\end{equation}
  
The above discussion relates the (multi) monopole mass to a set of topological charges which are specific
for the case of confined monopoles. There is a more generic set of topological charges which can be assigned to 
Coulomb (unconfined) and Higgs (confined) monopoles. The 
mass is determined by the behavior of the fields at the boundary at infinity which
we denote by $\del_\infty$. For the monopoles in the two different phases the boundary is of 
the form
\begin{equation}
  \label{boundary}
  \del_\infty = \left\{
                 \begin{array}{ll}
                        S^2_{\infty} &\ \ldots \mathrm{Coulomb}\\
                        \mathfrak{D}_+ \cup \mathfrak{D}_- \cup \mathfrak{Z}_\infty &\ \ldots \mathrm{Higgs}
                  \end{array} \right. \, ,
\end{equation}
where the first line denotes the sphere at infinity for the asymptotic Coulomb monopoles and for the 
confined monopole the boundary $\mathfrak{Z}_\infty$ gives no contributions (\ref{Mcl}). At the boundary
$\del_\infty$ the fields commute, i.e. $[\ca{B}_k\, ,\, \varphi]_{\del_\infty} =0$, see (\ref{asym3}) and 
\cite{Goddard:1976qe, Weinberg:1979zt} for the Coulomb monopole. 
The Cartan subalgebra can be therefore chosen such that it contains both asymptotic fields. 

The scalar field at the boundary is 
$\ \sqrt{2}\varphi_{\del_\infty}=\sqrt{2}\phi_0 = \bs{h} \cdot \bs{H} +O(r^{-1})$, 
where $\bs{H}=(H_i)$ are the Cartan generators and $\bs{h}$ is a constant 
vector\footnote{The approximation to order $O(r^{-1})$, $r$ being the spherical radial coordinate, 
applies to Coulomb monopoles \cite{Nitta:2010nd}
and is the maximum necessary information needed. For the confined monopoles the asymptotic value is 
reached with exponential precision
at the relevant boundary $\mathfrak{D}_\pm$ (\ref{asym3}).}. The square root $\sqrt{2}$ is a convention 
to suite the definition (\ref{massop}).
The asymptotic chromo-magnetic fields are of the form:  
\begin{align}
  \label{Bb}
      \mathrm{Coulomb:}&\quad   
      \ca{B}_{\del_{\infty}} \ \rightarrow\quad  \ca{B}_r = \frac{1}{4\pi r^2}\ \bs{g} \cdot \bs{H} + O(r^{-3})\, ,
       \nonumber\\[5pt]
       \mathrm{Higgs:}&\quad  
        \ca{B}_{\del_{\infty}}\  \rightarrow \quad \ca{B}_{3}^\pm = \frac{1}{ r}\del_r\ca{A}^\pm_\theta
         \quad \textrm{at} \ \  \mathfrak{D}_\pm\, .
\end{align}
In the Coulomb case $r$ denotes the spherical radial coordinate and 
$\bs{g}$ is a constant vector \cite{Coleman:1982cx} called 
``magnetic weight'' \cite{Goddard:1976qe}. The form at the boundary 
for the confined monopole follows from (\ref{asym3}), where in this case $r$ is
now the cylindrical radial coordinates. The gauge fields $\ca{A}^\pm_\theta$ lie also in 
the Cartan subalgebra and we associate to them a magnetic weight via their 
constant values at the boundary of the discs $\mathfrak{D}_\pm$ (\ref{asym2}) as 
follows: 
$\ca{A}_\theta|_{\mathfrak{\del D}_{+}} - \ca{A}_\theta|_{\mathfrak{\del D}_{-}} =: \frac{1}{2\pi}\ \bs{g}\cdot\bs{H}$.
We note that the $U(1)$-part of the $U(N)$ gauge field drops out of this definition, as it does in 
(\ref{ktop}). 

The magnetic weight $\bs{g}$ is  associated  
with the non-abelian magnetic flux through the surface at infinity,
\begin{equation}
  \label{flux}
  F_{\infty}\  =\ \int_{\del_\infty}\!\!\!d\sigma^k\,\ca{B}_k  \ =\  
    \frac{1}{4\pi}\int_{S^2_{\infty}} \bs{g} \cdot \bs{H} \ =\  
      \int_{\mathfrak{D}_{+} - \mathfrak{D}_{-}}\hspace{-8mm} \ca{B}_{3}
    \ =\  \bs{g} \cdot \bs{H}\, ,
\end{equation}
where we have assumed that the weights coincide for the Coulomb and Higgs monopole. Note that 
in both cases the flux $F_\infty$ contains no contribution from the $U(1)$ factor (for Coulomb
monopoles it decouples from the beginning). Thus if working with the $U(N)$ Cartan subalgebra,
see appendix \ref{CW}, the weights have to satisfy in both cases $\sum_i g_i=0$.  

Thus with every confined multimonopole in the Higgs phase one can associate a Coulomb 
multimonopole by identifying their magnetic weights and thus their total flux 
to infinity. They have then equal masses:
\begin{equation}
  \label{MCH}
   \bs{g}_{\mathrm{h}}\ =\  \bs{g}_{\mathrm{c}}\ =:\  \bs{g} \quad\quad \Rightarrow\quad\quad
    M_{\mathrm{h}}^{\mathrm{mon}} \ =\   M_{\mathrm{c}}^{\mathrm{mon}}\  = \ \frac{1}{g^2}\ \bs{h}\cdot \bs{g}\, ,
\end{equation}
where we assumed that for both cases the vacua $\phi_0$ are identical, for example (\ref{vac1}),
which gives $\bs{h}=(h_i=-2m_i)$. 
The first term in the energy (\ref{Ecl}), which is proportional to the spatial extent of the system
and is present only for confined monopoles, carries the infinite energy of the confinement transition 
and is exclusively given by the $U(1)$ factor of the gauge group.

In \cite{Goddard:1976qe} it was shown for Coulomb monopoles that the magnetic weight has to satisfy
a quantization condition in terms of the dual simple roots
\begin{equation}
  \label{simpl}
  \bs{g}\ =\ 4\pi \sum_{a=1}^{N-1} \mathfrak{n}_a\ \bs{\beta}^\ast_{(a)} \quad \textrm{with}\quad 
     \mathfrak{n}_a\in \mathbb{Z}
     \quad \textrm{and}\quad \bs{\beta}^\ast_{(a)}\ =\ 
     \frac{\bs{\beta}_{(a)}}{\bs{\beta}^2_{(a)}}\ =\ \bs{\beta}_{(a)} \, ,
\end{equation}
where $\bs{\beta}_{(a)}$ are the simple roots, we refer to appendix \ref{CW} for our conventions, and 
the integers $\mathfrak{n}_a$ are the so called GNO charges. 
If the unbroken gauge group is non-abelian, as we assume in general, not all simple roots will 
have a non-vanishing scalar product with $\bs{h}$. However, we can choose it to be non-negative,
as it is the case for the vacuum (\ref{vac1}): $\bs{h}\cdot \bs{\beta}_{(a)} = \sqrt{2}\,(m_{a+1}-m_{a})\geq 0$.
The GNO charges are separated accordingly into topological and non-topological charges \cite{Weinberg:1979zt}:
\begin{equation}
  \label{top}
  \bs{h}\cdot \bs{g} \ =\ \ 4\pi \sum_{a=1}^{N-1} \mathfrak{n}_a\ \bs{h}\cdot \bs{\beta}_{(a)} 
                      \ = \ 4\pi \sum_{t} \mathfrak{n}_t\ \bs{h}\cdot \bs{\beta}_{(t)}\, ,  
\end{equation}
where $\bs{\beta}_{(t)}$ are those simple roots that have a non-vanishing scalar product with $\bs{h}$.
For the vacuum $\phi_0$ (\ref{vac1}) with $q$ groups of $n_r$ degenerate masses, see (\ref{HCF}), 
there are $q-1$ topological charges and associated simple roots:
\begin{equation}
  \label{toproot}
    (\mathfrak{n}_t, \bs{\beta}_{(t)}) =  (\mathfrak{n}_{a_r}, \bs{\beta}_{(a_r)}) \quad \textrm{with} \quad 
     a_r = \sum_{s=1}^r n_s\ ,\ \  r=1,\ldots,q-1\, .
\end{equation}
Inserting the total flux relation  (\ref{flux}) into the definition 
of the principal topological charges for the confined monopole (\ref{ktop}) one finds
a relation with the topological GNO charges if one assumes that the magnetic weight for
the confined monopole is of the form (\ref{simpl}), or equivalently that the masses 
coincide\footnote{By the direct comparison of the masses \cite{Nitta:2010nd} arrived at 
similar relations, though there seem to be some typos in \cite{Nitta:2010nd}. To keep the formulas
compact we set the non-existing topological GNO charges $\mathfrak{n}_{a_0},  \mathfrak{n}_{a_N}$ 
(\ref{toproot}) equal to zero.} (\ref{MCH}):  
\begin{equation}
  \label{gnokvor}
  \Delta k^{\mathrm{vor}}_r = \mathfrak{n}_{a_r} - \mathfrak{n}_{a_{r-1}}\quad \Leftrightarrow\quad 
      \mathfrak{n}_{a_r} = \sum_{s=1}^r\Delta k_s^{\mathrm{vor}}\, , \quad 
      \mathfrak{n}^{\mathrm{tot}} := \sum_{r=1}^{q-1} \mathfrak{n}_{a_r} =
      -\sum_{r=1}^{q} r\,\Delta k_r^{\mathrm{vor}}\, ,
\end{equation}
where the last relation gives the total magnetic charge. This allows one to associate to every
confined multimonopole a Coulomb multimonopole, and vice versa, in a unique way. For example,
a single monopole with one GNO charge, $\mathfrak{n}_{a_{\hat{r}}}=1$ for $r=\hat{r}$ and otherwise zero,
corresponds to the vortex charges
$\Delta k^{\mathrm{vor}}_{\hat{r}} = 1 = - \Delta k^{\mathrm{vor}}_{\hat{r}+1}$. And a confined monopole with 
minimum set of vortex charges $\Delta k^{\mathrm{vor}}_{\hat{r}} = 1 = - \Delta k^{\mathrm{vor}}_{\hat{r}+\hat{s}}$ 
that obey (\ref{ktop}), correspond to $\hat{s}$ unit-GNO charges 
$ \mathfrak{n}_{a_{\hat{r}}}=\ \dots \mathfrak{n}_{a_{\hat{r}+\hat{s}-1}} =1$ with total magnetic charge 
$ \mathfrak{n}^{\mathrm{tot}} = \hat{s}$. Hence, for a confined monopole with unit total magnetic 
charge the confining vortices have to sit in neighboring group factors of the unbroken 
group (\ref{HCF}). This fits the picture that the confined monopoles correspond to kinks that 
connect neighboring vacua in $CP^n$ models, as it was recently confirmed also at the quantum level 
\cite{Burke:2011mw}.

\subsubsection*{Index and spectral density:}

The Index function $\ca{I}(M)$ (\ref{indres}) has two competing contributions, one from the adjoint 
sector and one from the fundamental sector. The contribution from the fundamental  sector, see (\ref{massop}),
is easily obtained with the explicit matrix realization (\ref{vac1}):
\begin{equation}
  \label{If}
  \ca{I}^{\mathrm{f}}(M)\ =\ \frac{1}{2}\, \sum_{i=1}^{N_f}\, \sum_{r=1}^{q}\
       \frac{\sqrt{2}(m_i-m_r)}{\sqrt{2(m_i-m_r)^2 + M^2}}\ \Delta k^{\mathrm{vor}}_r\ , 
\end{equation}
where we used the definition (\ref{ktop}). Note that in the limit $M\rightarrow 0$ the sum for  $i>N$ 
reduces to\footnote{We recall the ordering $m_1\leq\ldots\leq m_{N_f}$ of the masses, see below (\ref{lsur}).} 
$\sim \sum_{r=1}^q \Delta k^{\mathrm{vor}}_r =0$ by (\ref{ktop}). Hence, there are no 
contributions to the actual index $\ca{I}(0)$ from $i>N$, but there will be for the spectral density in 
(\ref{ind2}). 

The contribution to $\ca{I}(M)$ from the adjoint sector  is obtained by using the relations of 
appendix \ref{CW}. In particular, with $\sqrt{2}\phi_0^{\mathrm{a}} = \bs{h}\cdot\bs{H}^{\mathrm{ad}}$ (\ref{Had}),
one has for the adjoint corner in (\ref{massop}), 
\begin{equation}
  \label{mad}
   \left[ \frac{\ca{M}}{\sqrt{\ca{M}^2+M^2}}\right]^{\mathrm{a}} = 
      \begin{bmatrix} 0 & \cdot\cdot \\ : & 0 & i A\\
                           & - iA& 0\end{bmatrix}\, , \quad \textrm{with}\quad 
             [A] = \left[\frac{(\bs{h}\cdot\bs{\alpha})}{\sqrt{(\bs{h}\cdot\bs{\alpha})^2 + M^2}}\right]\, ,
\end{equation}
where $A$ is the diagonal matrix where each entry (\ref{mad}) corresponds to one of the positive 
roots $\bs{\alpha}$, see (\ref{Had}) and (\ref{roots}). The integration in (\ref{indres})
for the adjoint field $\ca{B}_3^{\mathrm{a}}$ gives the non-abelian flux (\ref{flux}) in the adjoint 
representation, i.e. $\bs{g} \cdot \bs{H}^{\mathrm{ad}}$, which is of the same form as (\ref{mad})
where the matrix elements of $A$ are now given by $(\bs{g} \cdot \bs{\alpha})$. The adjoint contribution 
to the index function $\ca{I}(M)$ (\ref{indres}) is then:
\begin{equation}
  \label{Ia}
  \ca{I}^{\mathrm{a}}(M)\ =\ -\frac{1}{2\pi}\, \sum_{\bs{\alpha}\,\in\,\Sigma^+} 
    \frac{(\bs{h}\cdot\bs{\alpha})\,(\bs{g} \cdot \bs{\alpha})}{\sqrt{(\bs{h}\cdot\bs{\alpha})^2 + M^2}}\, ,
\end{equation}
where the sum runs over all positive roots $\bs{\alpha}_{m<n}$, see below (\ref{roots}). Before elaborating 
on this expression we note that it can be written in a similar form as the fundamental contribution. One has,
\begin{equation}
  \label{af1}
  \bs{h}\cdot\bs{\alpha}_{k\ell} = \sqrt{2}\,(m_k - m_\ell)\ ,\quad 
  \bs{g} \cdot \bs{\alpha}_{k\ell} = 2\pi\, [\,(\mathfrak{n}_k - \mathfrak{n}_{k-1}) 
                          - (\mathfrak{n}_\ell - \mathfrak{n}_{\ell-1})\,]\, ,
\end{equation}
where we used the components of $\bs{h}$ given below (\ref{MCH}) and (\ref{simpl}). In the second
expression appear all GNO charges, also the non-topological ones. Inserting these expressions one 
finds
\begin{equation}
  \label{af2}
     \ca{I}^{\mathrm{a}}(M)\ =\  - \sum_{i=1}^{N}\, \sum_{r=1}^{q}\
       \frac{\sqrt{2}(m_i-m_r)}{\sqrt{2(m_i-m_r)^2 + M^2}}\ \Delta k^{\mathrm{vor}}_r\ , 
\end{equation}
where we have used that 
${\displaystyle{\sum_{s = a_{r-1}+1}^{a_r}}} (\mathfrak{n}_s - \mathfrak{n}_{s-1}) = \mathfrak{n}_{a_r} - \mathfrak{n}_{a_{r-1}} = 
\Delta k_r^{\mathrm{vor}}$. This is except for the difference in range of the first summation, which however 
does affect the index $\ca{I}(0)$ itself,  $(-2)$ times the fundamental contribution. 

The total index function $\ca{I}(M)$ can therefore be written as
\begin{equation}
  \label{indtot1}
       \ca{I}(M)\ =\  - \frac{1}{2}\, \Big{(}\, 2\sum_{i=1}^N - \sum_{i=1}^{\Nf}\, \Big{)}\, \sum_{r=1}^{q}\
       \frac{\sqrt{2}(m_i-m_r)}{\sqrt{2(m_i-m_r)^2 + M^2}}\ \Delta k^{\mathrm{vor}}_r\, .
\end{equation}
This is the form that was used in \cite{Burke:2011mw} to extract the spectral density according (\ref{ind2})
by a Laplace transformation:
\begin{equation}
  \label{specdens}
   \int_{\omega}\, \Delta \rho\ f(\omega^2) \ = \  
      \Big{(}\, 2\sum_{i=1}^N - \sum_{i=1}^{\Nf}\, \Big{)}\, \sum_{r=1}^{q}\ \int\,\frac{dk}{2\pi}\,
       \frac{- \sqrt{2}(m_i-m_r)}{k^2+ 2(m_i-m_r)^2 }\ \Delta k^{\mathrm{vor}}_r \ f(\omega^2)
    \, ,
\end{equation}
with $\omega^2= k^2+ 2(m_i-m_r)^2$ and  $f(\omega^2)$ an arbitrary function. This spectral density 
determines the full perturbative quantum energies of multiple confined monopoles and in addition 
an anomaly in the associated central charge, notably the magnetic charge in (\ref{Ecl}) \cite{Burke:2011mw}.  
As can be seen from (\ref{indtot1}), the fundamental and the adjoint sector compete in their  
contribution 
 to the index. However, this relative sign in the two sums of  (\ref{indtot1}) is 
of utmost importance in the quantum theory since it produces in the presence of confined monopoles
the $\beta$-function coefficient for the coupling constant renormalization. We refer to \cite{Burke:2011mw}
for further details. 

We can now turn to the computation of the index and its relation to the above discussed topological charges.
For this we consider the expression (\ref{Ia}). As already mentioned, the terms with $i>N$ 
in $\ca{I}^{\mathrm{f}}$ (\ref{If}) do not contribute for $M=0$ and thus up to a factor it is identical
to the adjoint contribution (\ref{Ia}). Concretely one has,
\begin{equation}
  \label{indaftot}
  \ca{I}\ :=\ \ca{I}(0) \ = \  
    \ca{I}^{\mathrm{a}}(0) + \ca{I}^{\mathrm{f}}(0)\  = \ \frac{1}{2}\ \ca{I}^{\mathrm{a}}(0)\, , 
\end{equation}
where  $\ca{I} = \mathrm{Ind}(L) = n_{L^\dagger L}^0 - n_{L L^\dagger}^0\,$ (\ref{ind1}). Thus the fundamental 
contribution halves the index of the adjoint sector. It is worth mentioning 
that the adjoint index (\ref{Ia}) for the confined monopoles is exactly the same\footnote{The overall 
minus sign is a convention that we inherited here from \cite{Burke:2011mw}.} as 
for the associated Coulomb monopoles \cite{Weinberg:1979zt}, which have the same magnetic weight $\bs{g}$.
It is difficult to say at this point what the halving of the index for the confined monopoles means. As 
it was discussed above, neither $L$ nor $L^\dagger$ are strictly positive (\ref{Lnorm}), and these
fluctuation operators describe the fluctuations of the full second order field equations and not just 
the $\frac{1}{4}$ BPS equations (\ref{BPS}). However, for so far in the literature considered $\frac{1}{2}$ BPS 
equations these two sorts of fluctuation operators were identical. We address this question in more
detail in the next section. 

In the evaluation of $\ca{I}^{\mathrm{a}}(0)$ one has to account for the fact that the factors 
$(\bs{h}\cdot\bs{\alpha})$ are non-negative, but that they cancel only if they are non-zero, 
otherwise there is no contribution from the respective root. Therefore the sum is restricted
to those positive roots $\bs{\alpha}_{m<n}$ for which $(\bs{h}\cdot\bs{\alpha})\neq 0$:
\begin{align}
  \label{controot}
     \ca{I}^{\mathrm{a}}(0)\ =& \ -\frac{1}{2\pi}\, \sum_{\bs{\alpha}\,\in\,\Sigma^+} \!\!\!{}' \
    (\bs{g} \cdot \bs{\alpha}) \ =\  \sum_{r=1}^{q-1}\ \  \sum_{{m =}\atop {a_{r-1} +1}}^{a_r}\ \ \sum_{n = a_r + 1}^N
     (\bs{g} \cdot \bs{\alpha}_{mn}) \nonumber\\[5pt]
     =& \ - \sum_{r=1}^{q-1}\, (n_{r+1} + n_{r})\, \mathfrak{n}_{a_r}\  = \ 
    \sum_{r=1}^{q}\, (a_{r} + a_{r-1})\, \Delta k_r^{\mathrm{vor}}\ , 
\end{align}
where the prime in the first sum indicates the discussed restriction. The function $a_r$ was defined 
in (\ref{toproot}) and $n_r$ are again the number of degenerated masses in the $r=1,\ldots q$ groups. 
The last line gives the index in terms of the two different topological charges that were introduced above.
For the case of abelian monopoles, i.e. maximal breaking of the gauge group ($n_r = 1$ for 
all $r = 1,\ldots q=N$), (\ref{controot})  reduces to the known result  \cite{Weinberg:1979zt},
\begin{equation}
  \label{indab}
  \ca{I}^{\mathrm{a}}(0)\ \ =\  -2\, \sum_{r=1}^{q-1}\, \mathfrak{n}_{a_r}\  =  -2\, \mathfrak{n}^{\mathrm{tot}}\, ,
\end{equation}
where $\mathfrak{n}^{\mathrm{tot}}$ is the total magnetic charge (\ref{gnokvor}). 

In \cite{Nitta:2010nd} some examples of the (framed) moduli space of Coulomb and confined monopoles were
explicitly constructed via the rational map and moduli matrix construction, and it was shown that they are 
identical. These moduli spaces are,
\begin{align}
  \label{mspaces}
      \ca{M}_{(\mathfrak{n}_{a_{\hat{r}}}=1)} = 
         \mathbb{R}^3\times S^1\ltimes(\mathbb{C}P^{n_{\hat{r}}-1}\times \mathbb{C}P^{n_{\hat{r}+1}-1})\, ,
      \quad \ca{M}_{(1,1)\mathrm{sep}} = (\mathbb{C} \times \mathbb{C}^*)^2\, .
\end{align}
The first one describes the moduli space of a Coulomb/Higgs monopole in the generic symmetry breaking 
pattern (\ref{HCF}) with unit charge $\mathfrak{n}_{a_{\hat{r}}} =1$. The second 
moduli space refers to a charge two Coulomb monopole in the completely broken phase of $SU(3)$ or, 
respectively, to a charge two confined monopole with symmetry breaking pattern $U(3)\rightarrow U(1)^3$
with $\mathfrak{n}_{1}=\mathfrak{n}_{2}=1$. The index $\mathrm{sep}$ indicates that the two monopoles
are confined by independent flux tubes, in contrast two freely moving monopoles aligned on a single 
confining vortex\footnote{\label{ftalign}In \cite{Nitta:2010nd} also the situation of two aligned monopoles 
is considered. The moduli space is of dimension six in this case. This does not fit in the 
the counting that we will propose in a moment. Apparently, the boundary conditions formulated in 
section \ref{secasym} do not account for such solutions where the monopoles on a given flux tube 
can be arbitrarily separated from each other, but describe only multimonopoles which are close or on top 
of each other if they are situated in the same confining vortex.} 
(in figure \ref{fig} this corresponds 
to the situation with one ``dot'' of charge one at each flux tube). A further example that is discussed
in \cite{Nitta:2010nd} is the charge two monopole in minimal breaking $SU(3)\rightarrow SU(2)\times U(1)$
or  $U(3)\rightarrow U(2)\times U(1)$ for the Coulomb and Higgs case respectively. 

In all these explicit cases one finds for the (real) dimension of the moduli space
\begin{equation}
  \label{dim}
  \mathrm{dim}(\,\ca{M}_{\mathrm{mod}}^{\mathrm{clb}}\,) \ = \ \mathrm{dim}(\,\ca{M}_{\mathrm{mod}}^{\mathrm{hgs}}\,) 
   \ = \
    2 |\,\ca{I}^{\mathrm{a}}(0)\,| =  4 |\,\ca{I}\,|\, , 
\end{equation}
with $\ca{I}^{\mathrm{a}}(0)$ given by (\ref{controot}).

We are thus led to the conjecture, that in general the index $\ca{I}$ (\ref{indaftot}) that we computed here, 
gives a quarter of the dimension of the moduli space of non-abelian confined multimonopoles, where 
on each separated flux tube the monopoles are confined to be close or on top of each other, see footnote
\ref{ftalign}.

\subsection{Parameter counting}

We discuss now the relation of the index which was computed in the previous section to
the dimension of the moduli space, given by the zero mode fluctuations around the BPS equations (\ref{BPS}). 

The operators (\ref{LLd}) were defined via the expansion of the Lagrangian  around 
solutions of the BPS equations (\ref{BPS}), and it is the supersymmetry structure, i.e.\ 
when taking into account the fermionic fluctuations, that determined how the 
fluctuations (\ref{W}) were organized and thus the form of the fluctuation operators 
(\ref{LLd}). We have to refer to \cite{Burke:2011mw} for details. One central result of 
this investigation is the spectral density (\ref{specdens}), obtained from the 
index function $\ca{I}(M)$ (\ref{indtot1}). 
However, for the evaluation of the index $\ca{I}=\ca{I}(0)$ itself we showed that 
the trivial fluctuation with $i>N$, see (\ref{W}), do not contribute. Therefore for the following 
comparison one has to consider only  the first $N$ flavors, or equivalently set
$N_f = N$.

\subsubsection*{BPS fluctuations}

Even for $N_f = N$ the fluctuations (\ref{W}) still contain the trivial fluctuations 
$s_{i2}$ which do not appear in the fluctuations of the BPS equations. The fluctuation 
equations for the BPS equations are obtained by inserting an expansion as given in (\ref{fluc}) 
for the classical fields in (\ref{BPS}) itself. However, the hermiticity condition 
(\ref{bkg}) for BPS solutions implies $a_5 = 0$ (it was irrelevant also before), 
and of course $s_{i2}$  is not present and we call the relevant fluctuation  
$s_{i1}=s_{i}$ in the following. 
In addition one has to implement the gauge fixing condition on the 
fluctuations. The gauge used in \cite{Burke:2011mw} to obtain the structure (\ref{l2}), amounts
to imposing the condition,
\begin{equation}
  \label{gfix}
  \ca{D}_{\ww k} a_{\ww k} + \fr{i g^2}{2} 
      \big{(} s_{i}\otimes\bar{\Sigma}_{i} - \Sigma_{i}\otimes \bar{s}_{i} \big{)} = 0\,,
\end{equation}
where the notation $\ww k$ was introduced in (\ref{euk}). The resulting fluctuation equations 
are given by,
\begin{align}
  \label{bpsfluc}
      \ca{E}_{(1)} &\ =\ \ca{D}_{\bar{z}} a_{z} + \ca{D}_{\bar{w}} a_{w} + i g^2 s_i\otimes \bar{\Sigma}_i\ =\ 0
     \, , \nonumber\\[4pt]
     \ca{E}_{(2)} &\ =\ \ca{D}_{z} s_i - i a_z \Sigma_i \ =\ 0   \, , \hspace{8mm}
     \ca{E}_{(2')} \ =\ \ca{D}_{w} s_i - i a_w \Sigma_i \ =\ 0  \, ,  \nonumber\\[4pt]
       \ca{E}_{(3)} &\ =\ \ca{D}_{w} a_z - \ca{D}_{z} a_w \ =\ 0  \, .
\end{align}
Hereby is $\ca{E}_{(1)}$ the fluctuation equation for $k=1$ in (\ref{BPS}) and $\ca{E}_{(3)}$ 
is a complex combination of the $k=1$ and $k=2$ fluctuation equations. As mentioned below (\ref{BPS}),
the $k=1,2$ BPS equations are the integrability conditions for the two last ones. Similarly one finds that 
fluctuation equations $\ca{E}_{(2)}, \ca{E}_{(2')}$ imply the fluctuation equation $ \ca{E}_{(3)}$: 
$\ca{D}_w \ca{E}_{(2)} - \ca{D}_z \ca{E}_{(2')} = i\, \ca{E}_{(3)} \Sigma_i$. 

Therefore, only the first 
three equations in (\ref{bpsfluc}) are independent and define the fluctuation operator:
\begin{equation}
  \label{bpsLd}
    \widehat{W} = [\ \fr{1}{g}\, a_w\, ,\ \fr{1}{g}\, a_z\, ,\ s_{i}\, ]^T \quad \Rightarrow
   \quad \hat{L}^\dagger = 
    \begin{bmatrix}\, - \ca{D}_{\bar w}^{\mathrm{a}} & \, - \ca{D}_{\bar z}^{\mathrm{a}}&\,  
                           -ig \bar{\Sigma}_i^{\mathrm{r}}\, \\[3pt]
                         \, ig \Sigma_i^{\mathrm{r}}& 0&\,  - \ca{D}_{w}^{\mathrm{f}}\,  \\[3pt]
                        0& ig\Sigma_i^{\mathrm{r}}& - \ca{D}_{z}^{\mathrm{f}}\, \end{bmatrix}\, ,
\end{equation}
where the fluctuation field $\widehat{W}$ is the same as in (\ref{W}) with the last entry, the trivial 
fluctuation $s_{i2}$, deleted. The ordering and notation for the fluctuation operator will become clear 
in a moment. The operator adjoint to the one in (\ref{bpsLd}) reads as,
\begin{equation}
  \label{bpsL}
  \hat{L} = 
    \begin{bmatrix}\,  \ca{D}_{ w}^{\mathrm{a}} & \, - ig \bar{\Sigma}_i^{\mathrm{r}} & \, 0 
                          \, \\[3pt]
                         \, \ca{D}_{z}^{\mathrm{a}} & 0&\,  - ig\bar{\Sigma}_i^{\mathrm{r}} \,  \\[3pt]
 \,  ig \Sigma_i^{\mathrm{r}} & \,  \ca{D}_{\bar{w}}^{\mathrm{f}} &  \ca{D}_{\bar{z}}^{\mathrm{f}}\, \end{bmatrix}\, .
\end{equation}
The two operators act in the spaces 
$\hat{L}^\dagger:\left[{X^\alpha\atop y_i}\right] \rightarrow \left[{X'\atop y'^\alpha_i}\right]$
and $\hat{L}:\left[{X\atop y^\alpha_i}\right] \rightarrow \left[{X'^\alpha\atop y'_i}\right]$, with 
$\alpha=1,2$. We used here the same notation as beneath (\ref{euk}). These spaces are all of dimension 
$3N^2 = 2\, d_{\mathrm{ad}} + N\times d_{\mathrm{fun}} = 
d_{\mathrm{ad}} + 2 N\times d_{\mathrm{fun}}$, where $d_{\mathrm{ad}}, d_{\mathrm{fun}}$ denote the dimension 
of the respective representations of $U(N)$. 

Contrary to the previous situation, one of the two operators is now strictly positive,
\begin{equation}
  \label{Lbpsnorm}
  \left\| \hat{L}\,  {\textstyle{\left[{X\atop y^\alpha_i}\right]}}  \right\|^2 \ = \  
      \left\| \ldots \right\|^2
      +g^2 \big{(} \left\|y^\alpha_i\otimes \bar\Sigma_i \right\|^2 
          +  \left\| X \Sigma_i \right\|^2\big{)} \overset{!}{=}0\ \ \  \Rightarrow \ \ \
    {\textstyle{\left[{X\atop y^\alpha_i}\right]}} = 0\, ,
\end{equation}
and therefore the index $\mathrm{ind}(\hat{L}^\dagger)$ computes the number of zero modes of $\hat{L}^\dagger$, 
i.e.\ the (complex) dimension of the moduli space of solutions of the BPS equations (\ref{BPS}). However,
there is a relation to the previously considered operators (\ref{LLd}). Using the explicit matrices 
(\ref{euk}), one finds that by deleting certain rows and columns from the operators (\ref{LLd}) that,
\begin{equation}
  \label{LbpsL}
  \hat{L}^\dagger = L^\dagger |_{(2,4)}\ \, , \qquad \hat{L} = L|_{(4,2)}\, ,
\end{equation}
where we have indicated which $(\mathrm{row}, \mathrm{column})$ has to be deleted 
to give the BPS fluctuation operators (\ref{bpsLd}), (\ref{bpsL}). 

For the first operator $L^\dagger$, the deletion described in (\ref{LbpsL}), amounts to 
setting the last component in $W$ (\ref{W}), i.e. the trivial fluctuation $s_{i2}$, to zero. 
The second row of $L$ reduces then to
the integrability condition $\ca{E}_{(3)}$ (\ref{bpsfluc}), which is implied by the 
other equations for zero modes. The question is if this is consistent, 
i.e.\  if all zero modes of $L^\dagger$ are of this form. Inspection of the product $LL^\dagger$, which has 
the same zero modes as $L^\dagger$, shows that with (\ref{ddbps}) the last component of $W$ decouples. The 
zero mode equation of $L^\dagger$ for this decoupled mode implies $s_{i2}\otimes\bar{\Sigma}_i = 0$ so that
there are no such nontrivial zero modes. Consequently, the zero modes of $L^\dagger$ and $\hat{L}^\dagger$
are identical, i.e.\
\begin{equation}
  \label{llzm}
     L^\dagger\, W =0 \ \ \Rightarrow \ \ W = [ X^1,\, X^2,\, y_i,\, 0]^T
       \ \ \Rightarrow \ \ n^0(L^\dagger) = n^0(\hat{L}^\dagger)\, ,
\end{equation}
where $n^0$ denotes the respective number of zero modes.

The situation is different for the second set of operators in (\ref{LbpsL}). Inspection of the product
$L^\dagger L$ (\ref{LL}) shows that in this case the first component of the fluctuation 
field decouples and that the  zero mode equation of $L$ implies $X^1\Sigma_i=0$ and thus there is no such 
zero mode. Nontrivial zero modes are thus of the form $W = [ 0,\, X,\, y^1_i,\, y^2_i]^T$. Again, one
of the component zero mode equations of $L$ is implied by the other three via the integrability 
condition. The residual three equations can be encoded in an operator\footnote{In order to emphasize the 
similarities with $\hat{L}^\dagger$ we ordered the fluctuation field as 
$\hhat{W} = [ - y^2_i,\, y^1_i, \, X]^T$ for the definition of $\hhat{L}$.} $\hhat{L}$:
\begin{equation}
  \label{Lhh}
  L\, W = 0 \qquad \Rightarrow\qquad \hhat{L} = 
   \begin{bmatrix}\, - \ca{D}_{ w}^{\mathrm{f}} & \, - \ca{D}_{ z}^{\mathrm{f}}&\,  
                           -ig \Sigma_i^{\mathrm{r}}\, \\[3pt]
                         \, ig \bar{\Sigma}_i^{\mathrm{r}}& 0&\,  - \ca{D}_{\bar{w}}^{\mathrm{a}}\,  \\[3pt]
                        0& ig\bar{\Sigma}_i^{\mathrm{r}}& - \ca{D}_{\bar{z}}^{\mathrm{a}}\, \end{bmatrix}\, . 
\end{equation}

The deletion of row and column as given in (\ref{LbpsL}), on the other hand, amounts to setting  the 
second component in the fluctuation field to zero, instead of the first one, and in addition one needs
a nontrivial equation which is not implied by an integrability condition. Clearly the zero modes
of $L$ and $\hat{L}$ are different, the latter one does not have any (\ref{Lbpsnorm})  whereas 
$L$ is not necessarily positive (\ref{Lnorm}), and according to the conjecture (\ref{dim}) 
the number of zero modes is,
\begin{equation}
  \label{Lzero}
    n^0(L) = n^0(\hhat{L}) = \frac{1}{2}\,n^0(\hat{L}^\dagger)\, .
\end{equation}

The operators $\hat{L}^\dagger$ and  $\hhat{L}$ have a similar structure, with the roles of fundamental and 
adjoint representation interchanged, and a complex conjugation of the entries. One could try to find a 
$2:1$  map between the zero mode solutions of $\hat{L}^\dagger$ and  $\hhat{L}$ to proof the last equality in 
(\ref{Lzero}). Certainly one could also try to compute directly the index of $\hat{L}^\dagger$. In this 
regard we want to mention one difference to the considerations in the previous sections. The 
auxiliary Hamiltonian (\ref{H}) is now with (\ref{bpsLd}), (\ref{bpsL}) of the form,
\begin{equation}
  \label{Hhaux}
  \hat{H} = \hat{\Gamma}^i\del_i + \hat{K} \quad\textrm{with}\quad 
            \{\,\hat{\Gamma}^i, \hat{\Gamma}^j\}\neq 2\, \delta^{ij}\, ,
\end{equation}
i.e.\ the $\hat{\Gamma}^i$-matrices do not satisfy a Clifford algebra. A number of manipulations of the previous 
section need only that these matrices anti-commute with $\Gamma_5$, which is still the case, but many other 
details will be different. We have to leave both of the mentioned considerations for a separate 
investigation.  

\section{Summary and Conclusions}

In this paper we computed the index and the associated spectral density, which plays a central 
role in quantum computations \cite{Burke:2011mw}, for fluctuation 
operators which are defined via the Lagrangian in a confined multimonopole background. These 
confined monopole backgrounds describe asymptotically nontrivial field configurations. 
To this end it was necessary to generalize the standard index calculations of
\cite{Callias:1977kg,Weinberg:1979ma,Weinberg:1981eu} appropriately. 
The general strategy  for such computations is as follows:

We assume that the boundary of the space occupied by the 
background is the sum of two components, 
$\del\mathbb{M} = \del\mathbb{M}_{\mathrm{triv}} \cup  \del\mathbb{M}_{\mathrm{nontriv}}$, where on one 
component the field configurations are trivial but not on the other. Then i.) reformulate the index as 
a sum of an anomaly and a surface term using the standard techniques. ii.) The anomaly 
and the surface term on the boundary $\del\mathbb{M}_{\mathrm{triv}}$ can be computed in the 
limits as outlined in \cite{Callias:1977kg,Weinberg:1979ma,Weinberg:1981eu}. iii.) Rewrite the
nontrivial surface term on $\del\mathbb{M}_{\mathrm{nontriv}}$ as a generalized index, as in 
(\ref{dcur}). iv.) Fourier transform the derivative operators which transverse to
the boundary $\del\mathbb{M}_{\mathrm{nontriv}}$. v.) Find a proper factorization of the 
auxiliary Hamiltonian for the index on the boundary and reformulate this index as the sum of an anomaly 
and a surface term. If the fields are trivial at the boundary of  $\del\mathbb{M}_{\mathrm{nontriv}}$,
which has to be the case for the above splitting of $\del\mathbb{M}$ to be consistent, all terms
can be computed in the convenient limits. 

The resulting index for the confined monopoles is exclusively given by the anomaly 
$\ca{A}^{\mathrm{disc}}$ (\ref{indres}) on the 
discs $\mathfrak{D}_{\pm}$, which represent the boundary $\del\mathbb{M}_{\mathrm{nontriv}}$ in this case. 
Contrary to standard index calculations, this anomaly depends nontrivially on the IR regulator mass $M$ and 
thus leads to a non-vanishing spectral density (\ref{specdens}). 
We were able to express this index in terms of topological charges of the confined 
monopoles, which we also related to topological charges of an associated Coulomb monopole (\ref{controot}).
From some existing examples of the moduli space of confined monopoles we conjectured that 
the index presented  here  counts a quarter of the dimension of the moduli space (\ref{dim}). Finally we compared 
the fluctuation operators  defined via the Lagrangian with those of the BPS equations. However,
a detailed proof of the conjecture regarding the parameter counting has to be left for a separate investigation.
\\

%%%%%%%%%%%%%%%%%%%%%%%%%%%%%%%%%%%%%%%%%%%%%%%%%%%%%%%%%%%%%%%%%%%%%%

\noindent
{\bf Acknowledgments:} I thank A. Rebhan and D. Burke for many useful comments on the script.
This work is supported by the Agence Nationale de la Recherche (ANR).

\appendix

\section{Overall Anomaly}\label{anotot}

Here and for other proofs we will need some properties of the free massive Euclidean Green's
function, which we henceforth call ``propagator''.
In $D>1$ dimensions the Euclidean propagator is given by
\begin{align}
  \label{massprop}
  \Delta_\mu (x-y) :=\frac{1}{-\del^2+\mu^2}&\ = \int\,\frac{d^Dk}{(2\pi)^D}\,\frac{e^{i k(x-y)}}{k^2+\mu^2}
      \nonumber\\[5pt]
     &\ = \frac{1}{\sqrt{\pi}}\,\frac{\Gamma(\nu+\hal)}{(2\pi^2)^\nu}\
     \Big{(}\,\frac{\mu}{|x-y|}\,\Big{)}^\nu\, K_\nu(\,\mu|x-y|\,)\, , \quad \nu = \fr{D-2}{2}\, ,
\end{align}
where $K_\nu(z)$ is the modified Bessel function of 2'nd kind and for $D=2,3,4,\ldots$ one 
has $\nu=0,\hal,1,\ldots$ We note that $\Delta_\mu(\lambda z) = \lambda^{-2\nu}\Delta_{\lambda\mu}(z)$. 
The asymptotic behavior of the propagator is dominated by the exponential
decay of the Bessel function. The large mass limit is of the form
\begin{equation}
  \label{largmu}
  \Delta_\mu (x-y) \quad \overset{\mu\rightarrow\infty}{\longrightarrow}\quad 
    \frac{1}{\sqrt{2}}\, \frac{\Gamma(\nu+\hal)}{(2\pi^2)^\nu}\ \frac{\mu^{\nu-1/2}}{|x-y|^{\nu+1/2}}\
     e^{-\mu|x-y|} \ \big{[}\,1+O(\fr{1}{\mu})\,\big{]}\, .
\end{equation}
The folded product of $n+1$ propagators,
\begin{align}
  \label{enfold}
   \Delta_\mu^{(n+1)}(x-y):=&\ \int_{z_1,\ldots z_n} \Delta_\mu(x-z_1) 
                        \Delta_\mu(z_1-z_2) \ldots \Delta_\mu(z_n-y) \nonumber\\[5pt]
       = &\  \int\,\frac{d^Dk}{(2\pi)^D}\,\frac{e^{i k(x-y)}}{(k^2+\mu^2)^{n+1}}
       \ \ \overset{y\rightarrow x}{\longrightarrow}\ \
       \frac{1}{(4\pi)^{\nu+1}}\,\frac{\Gamma(n-\nu)}{\Gamma(n+1)}\ \mu^{2(\nu-n)}\ , 
\end{align}
is regular at $y=x$ for $n >\nu=0,\hal,1\ldots$ and scales as $\sim \mu^{2(\nu-n)}$ in this case. 
The insertion of $r=1,2,\ldots, n+1$ derivatives
$\del_{i_1}\ldots\del_{i_r}$ in the 
folded product (\ref{enfold}), 
distributing them arbitrarily while acting always on the first argument of the respective propagator,
\footnote{Due to translation invariance one
 may always assume that the derivatives act on the first argument of the propagators in  (\ref{enfold}).} 
gives 
\begin{align}
  \label{defold}
        \int\,\frac{d^Dk}{(2\pi)^D}\,\frac{i^r k_{i_1}\ldots k_{i_r}\,e^{i k(x-y)}}{(k^2+\mu^2)^{n+1}} 
         &\ =\ \del_{x^{i_1}}\ldots\del_{x^{i_r}}\Delta_\mu^{(n+1)}(x-y) \nonumber\\[3pt]
        &\overset{y\rightarrow x}{\longrightarrow}\ \frac{i^r}{(2\pi)^D}\int_{\Omega_{D-1}} 
    \hat{k}_{i_1}\ldots \hat{k}_{i_r} 
            \int_0^\infty \frac{dk\,k^{D-1+r}}{(k^2+\mu^2)^{n+1}}\, .
\end{align}
Regularity at $y=x$ is given for $2n-r > 2\nu=0,1,2\ldots$ and 
the resulting expression scales as  $\sim \mu^{2\nu-(2n-r)}$.
In this case the integration over the unit sphere $\Omega_{D-1}$ ($\hat{k}_i =k_i/k$) vanishes for 
odd $r$ by symmetric  integration.   
\\

\noindent
In computing the overall anomaly in (\ref{overan}) one has to evaluate the 
different contributions in the expansion (\ref{anex})  for large $\mu$ up to orders $O(\mu^{-3})$. To this end
one changes the integration\footnote{We follow here \cite{Callias:1977kg} with the difference that the 
insertions $\ca{O}$ contain also derivative operators.} variables to $z_i \rightarrow x+z_i/\mu$. The 
expansion (\ref{anex}) then reads as
\begin{equation}
  \label{anextrans}
  \mu^{2(\nu-n)} \sum_{n\geq 1} \int_{z_1,\ldots z_n} \hspace{-8mm}\Delta_1(-z_1) 
                        \ca{O}(x+\fr{z_1}{\mu})\Delta_1(z_1-z_2) \ldots \ca{O}(x+\fr{z_n}{\mu})
      \Delta_1(z_n+\mu(x-y))\, ,  
\end{equation}
where we used the scaling described below (\ref{massprop}) and $\nu = \hal$ here. 
The operator insertions (\ref{O}) 
are of the form
\begin{equation}
  \label{Otrans}
   \ca{O}(x+\fr{z_i}{\mu}) = C(x+\fr{z_i}{\mu}) + \mu\, K_{k}\,(x+\fr{z_i}{\mu})\,\del_{z_i^k}\, ,
\end{equation}
so that the $n$'th term in the expansion (\ref{anextrans}) for $y=x$ is of order $\sim \mu^{2\nu-(2n-r)}$ with $r$
being the number of derivatives of (\ref{Otrans}) that appear in the product. The expansion of the coefficient 
functions in (\ref{Otrans}) around $x$ gives further inverse powers of $\mu$, while the $z$-integrations 
for these terms are finite due to the exponential decay of the propagators. 

As mentioned in the main text, for any given order $n$ in the expansion 
(\ref{anextrans}) the term with exclusively derivative insertions vanishes under the trace with 
$\Gamma_5$. Thus in three dimensions ($2\nu =1$) only the terms with $(n=1,r=0)$ and $(n=2,r=1)$ survive
the $\mu\rightarrow\infty$ limit. Both terms  are regular\footnote{Only the $(n=1,r=1)$ term 
diverges at $y=x$ in the expansion (\ref{anextrans}), but as noted its trace with $\Gamma_5$ vanishes.} 
at $y=x$.

Noting that $\mathrm{Tr}\,\{\Gamma_5 K^2 \}=0$, see  (\ref{O}) and (\ref{g5}), 
the  $(n=1,r=0)$ contribution to the anomaly can be written as
\begin{align}
  \label{J1}
  \ca{J}^{\mathrm{ano}}_{(1)} (x,x) =&\ \lim_{\mu\rightarrow\infty} \mu\, \int_z  \mathrm{Tr} \{ 
     \Gamma_5 \Gamma^i\del_iK (x+\fr{z}{\mu})\}\, \Delta_1(-z)\, \Delta_1(z) \nonumber\\[4pt]
   &\ = \lim_{\mu\rightarrow \infty}\, 
        \left[\,\frac{\mu}{8\pi}\, \mathrm{Tr}\,\{\,\Gamma_5 \Gamma^i\del_iK(x)\,\}\right]\, .
\end{align}
In the second step we expanded the insertion around $x$ and used (\ref{enfold}) for $\nu=\hal$.
The order $\sim z/\mu$ term in the expansion of the insertion would survive the 
$\mu\rightarrow\infty$ limit but it vanishes by symmetry of the $z$-integration. 

For the $(n=2,r=1)$ term the coefficient functions of (\ref{Otrans}) can be directly evaluated 
at $x$. The resulting term is proportional to a folded product of three propagators 
with a single derivative insertion and thus vanishes by symmetric integration, see (\ref{defold}).
This leaves for the overall anomaly the contribution (\ref{J1}), which vanishes due to the trace 
with $\Gamma_5$, see (\ref{jan}).

\section{Disc anomaly and Surface term}\label{das}

As shown in  (\ref{sst}), the index function $\ca{I}(M)$ for confined monopoles reduces 
to a surface term and anomaly contribution on the discs at infinity $\mathfrak{D}_{\pm}$. We give here some 
details in the evaluation of these contributions. 

In what follows one needs the explicit form of the products  
$\ca{L}\ca{L}^\dagger$, $\ca{L}^\dagger\ca{L}$  on the discs $\mathfrak{D}_{\pm}$ 
(\ref{lld}). Both products contain an adjoint and fundamental ``mass term'' which can be seen in the 
diagonal elements of the original operators (\ref{LL}). With the behavior at the discs $\mathfrak{D}_{\pm}$
as described around (\ref{asym3}) the explicit matrix form of these terms reads as
\begin{align}
  \label{massdisc}
   \bar\Sigma_iT^AT^B\Sigma_i\,|_{\mathfrak{D}_{\pm}}&\, =\ \sum_{i=1}^N \, |\sigma_i^\pm|^2\, (T^AT^B)^i{}_i
     &\underset{r\rightarrow\infty}{\longrightarrow}&\qquad \frac{v^2}{2}\delta^{AB}
    \nonumber\\[5pt]
   \bar\Sigma_j\Sigma_i\,|_{\mathfrak{D}_{\pm}}&\, =\
      \mathrm{diag}(|\sigma_1^\pm|^2,\ldots,|\sigma_N^\pm|^2,0,\ldots,0)_{ij}
   &\underset{r\rightarrow\infty}{\longrightarrow}&\qquad v^2 I_{ij}^N\, 
\end{align}
where the $N_f\times N_f$ matrix  $I_{ij}^N$ has unit entries on the first $N$ diagonal elements and
is otherwise zero. The expressions on the boundary  $\del\mathfrak{D}_{\pm}$ of the discs are given
up to exponentially suppressed terms. The operators on the disc are then given by
\begin{align}
  \label{opsdisc}
    \ca{L}^\dagger\ca{L} = -\del_\alpha^2 + g^2v^2 I' -(b_\alpha\del_\alpha + c + \Delta c)
     \ , \
   \ca{L}\ca{L}^\dagger =  -\del_\alpha^2 + g^2v^2 I' -(b_\alpha\del_\alpha + c + \tilde{\Delta c}),   
\end{align}
so that they differ only in the last term. The ``identity'' 
$I'=\mathrm{diag}(\unit^{\mathrm{a}}, \unit^{\mathrm{f}}\,I^N_{ij})$ has zero 
entries for $i,j > N$. The respective terms are
\begin{align}
  \label{terms}
    b_\alpha = -2\,i \begin{bmatrix} \ca{A}_\alpha^{\mathrm{a}}&\\ & \delta_{ij} \ca{A}_\alpha^{\mathrm{f}}\end{bmatrix}
   \ , \qquad
 \Delta c =\begin{bmatrix} \ca{B}_3^{\mathrm{a}}&\\ & -\delta_{ij} \ca{B}_3^{\mathrm{f}} \end{bmatrix}
    \ , \qquad
   \tilde{\Delta c} =\begin{bmatrix} -\ca{B}_3^{\mathrm{a}}& \ast\\ 
             \ast^\dagger &  \delta_{ij} \ca{B}_3^{\mathrm{f}} \end{bmatrix}\, ,
\end{align}
 where the off-diagonal terms 
$\ast \sim  \ca{D}_z\bar\Sigma_j$ vanish exponentially for $r\rightarrow\infty$, see (\ref{asym2}). They will 
drop out in the following analysis. The same applies to $c$ in (\ref{opsdisc}), which is diagonal 
and contains the terms $\ca{D}_\alpha A_\alpha$, 
$(\bar\Sigma_iT^AT^B\Sigma_i\,|_{\mathfrak{D}_{\pm}} -\frac{v^2}{2}\delta^{AB})$ and 
$(\bar\Sigma_j\Sigma_i\,|_{\mathfrak{D}_{\pm}} - v^2 I_{ij}^N)$, all of which vanish exponentially 
for $r\rightarrow\infty$. Since also the chromo-magnetic field $\ca{B}_3$ vanishes exponentially
we note that the two operator products in (\ref{opsdisc}) coincide at the boundary 
$\del\mathfrak{D}_{\pm}$ with only $b_\alpha$ vanishing polynomial:
\begin{align}
  \label{discbound}
 \del\mathfrak{D}_{\pm}:\ 
  \ca{L}\ca{L}^\dagger\approx  \ca{L}^\dagger\ca{L} \approx (-\del_\alpha^2 + g^2v^2 I') -b_\alpha\del_\alpha\ ,\quad
 b_{\alpha}\del_\alpha \approx -\frac{2\,i}{r}
       \left [{\!\!\!\!\!\!\!\!\!\ca{A}^{\mathrm{a}}_\theta  \atop \quad\ \ca{A}^{\mathrm{f}}_\theta}\right] 
     \vep^{\alpha\beta}\hat{x}_\alpha\del_\beta\, ,
\end{align}
where $\ca{A}_\theta$ is constant and commutes with $\phi_0$, which can be therefore chosen to be in the 
Cartan subalgebra ($\hat{x}_\alpha$ is the unit-vector). The term $b_\alpha$ therefore vanishes only 
like $\sim \frac{1}{r}$ and contributes to the surface integral.  
\\

\noindent
{\bf Anomaly.} The anomaly contribution in (\ref{sst}), with the help of (\ref{jred2}), can be written as
\begin{equation}
  \label{a}
    \mathrm{a}(x) = \mathrm{Tr}\, \big{\{}\, \frac{\mu^2-M^2}{\ca{M}^2+M_p^2}\, 2\,\ca{M}\,
                     \Big{[} \frac{1}{\ca{L}^\dagger\ca{L}+\ca{M}^2+\mu_p^2} - 
                             \frac{1}{\ca{L}\ca{L}^\dagger+\ca{M}^2+\mu_p^2}\, \Big{]}\, \big{\}}|_{y=x} \, .
\end{equation}

As for the overall anomaly, discussed in appendix \ref{anotot}, we expand the Green's functions in (\ref{a}) 
for large $\mu$, see (\ref{anex}), keeping the terms that survive the limit $\mu\rightarrow\infty$ of (\ref{a}). The 
Green's function in question take with (\ref{opsdisc}) the form
\begin{align}
  \label{expand}
   \frac{1}{\ca{L}^\dagger\ca{L}+\ca{M}^2+\mu_p^2} = 
         \frac{1}{-\del^2 + (g^2v^2 I'+\ca{M}^2+\mu^2+p^2) -(b_\alpha\del_\alpha + c + \Delta c)}\ ,
\end{align}
and analogously for $\ca{L}\ca{L}^\dagger$, replacing $\Delta c$ with $\tilde{\Delta c}$.
The first term in the expansion is the propagator $\Delta_{\mathfrak{m}}(x-y)$, which is obtained
from the $D=2$ case of (\ref{massprop}) by replacing the mass $\mu$ by the mass-matrix
\begin{equation}
  \label{mMat}
  \mathfrak{m}^2 = g^2v^2 I'+\ca{M}^2+\mu^2+p^2\, .
\end{equation}
Thus the propagator inherits the matrix structure of 
the diagonal\footnote{Note that the entries of $\ca{M}$, see 
(\ref{massop}), live in 
the Cartan subalgebra so that the fundamental and adjoint entry of $\ca{M}^2$ are diagonal, see appendix \ref{CW}.} 
 matrix $\mathfrak{m}^2$. However, these contributions cancel between the two Green's functions in 
(\ref{a}). The residual terms in the expansion are of the form (\ref{anextrans})
with $\Delta_1 \rightarrow \Delta_{\mathfrak{m}/\mu}$ and
\begin{equation}
  \label{a2O}
  \ca{O}(x+\fr{z}{\mu}) = c\, (x+\fr{z}{\mu}) + \Delta c\,(x+\fr{z}{\mu})
          + \mu\, b_\alpha (x+\fr{z}{\mu})\, \partial_{z^\alpha}\, ,
\end{equation}
and the same for the second Green's function in (\ref{a}) with  $\Delta c$ replaced by $\tilde{\Delta c}$.

On the two-dimensional discs $\nu = 0$ and therefore each term in the expansion (\ref{anextrans})
is now regular at $y=x$, see below (\ref{defold}). Also, the $n$'th term in the expansion is now 
proportional to $\sim \mu^{-2n+r}$, $r$ being again the number of derivatives, and thus only 
the contributions $(n=1,r=0,1)$ and $(n=2,r=2)$ survive the limit $\mu\rightarrow\infty$ of (\ref{a}).
However, the contributions with the maximum number $r=n$ of derivatives are the same for both Green's functions
in (\ref{a}) and thus cancel each other. 

Therefore one is left with the single contribution $(n=1,r=0)$ to (\ref{a}):
\begin{align}
  \label{anres}
   \mathrm{a}(x) =&\  \mathrm{Tr}\, \big{\{}\, \frac{\mu^2-M^2}{\ca{M}^2+M_p^2}\, 2\,\ca{M}\
                     \mu^{-2}\int_z \Delta_{\mathfrak{m}/\mu}(-z)\ \big{[}
                         \Delta c - \tilde{\Delta c} \big{]}(x+\fr{z}{\mu})\
                    \Delta_{\mathfrak{m}/\mu} (z) \, \big{\}} \ldots \nonumber\\[9pt]
                =&\ \frac{1}{\pi}\  \mathrm{Tr}\, \Big{\{}\, \frac{\ca{M}}{\ca{M}^2+M_p^2}\ 
                      \frac{\mu^2-M^2}{g^2v^2 I'+\ca{M}^2+\mu^2+p^2}\ 
                       \begin{bmatrix} \ca{B}_3^{\mathrm{a}}&\\[-4pt] 
                           & - \delta_{ij} \ca{B}_3^{\mathrm{f}}\end{bmatrix} \Big{\}} 
               +O(\mu^{-1})\, .          
\end{align}
In the second step we evaluated the difference $[ \Delta c - \tilde{\Delta c}]$ at $x$ 
and used that at the discs $\mathfrak{D}_{\pm}$ the field  
$\ca{B}$ commutes with $\ca{M}$, and thus with the propagators $\Delta_{\mathfrak{m}/\mu}$,
see below (\ref{asym3}). In addition, we note that 
$\Delta_{\mathfrak{m}/\mu}^2(0) =\frac{1}{4\pi} \frac{\mu^2}{\mathfrak{m}^2}$, see (\ref{enfold}) with 
$\nu=0$.

The actual local anomaly contribution (\ref{sst}) is the integration over $p$ of $\mathrm{a}(x)$.
However, it commutes with the limit $\mu\rightarrow\infty$ which puts the second factor in the 
trace to one. 
This leaves for the  $p$-integration,
\begin{align}
  \label{pint}
   \int\frac{dp}{2\pi}\ \frac{\ca{M}}{(\ca{M} + M^2 + p^2)}
    = \frac{\ca{M}}{2\sqrt{\ca{M}^2+M^2}}\, ,
\end{align}
where we note that the integral is well defined. The possible zero eigen-values
of $\ca{M}^2$ drop out because of the factor $\ca{M}$ in the numerator. Therefore even for $M\rightarrow 0$, 
a limit that is eventually of interest, the integrand has no poles.  

We thus obtain for the disc anomaly contribution to the index function (\ref{discIM}), (\ref{sst}),
\begin{align}
  \label{discanomaly}
    \ca{A}^{\mathrm{disc}} = -\frac{1}{4\pi} \int_{\mathfrak{D}_{+} - \mathfrak{D}_{-}}\hspace{-4mm}
                          \mathrm{Tr}\, \Big{\{} \frac{\ca{M}}{\sqrt{\ca{M}^2+M^2}}
                 \begin{bmatrix} \ca{B}_3^{\mathrm{a}}&\\[-4pt] 
                           & - \delta_{ij} \ca{B}_3^{\mathrm{f}}\end{bmatrix} \Big{\}}  \, ,           
\end{align}
which is in fact $M$-dependent.
\\

\noindent
{\bf Surface term.} Contrary to the anomaly, the surface term is evaluated at the 
boundary $\del\mathfrak{D}_{\pm}$ of the discs. Using (\ref{jred2}) the current in (\ref{sst}) 
can be written as
\begin{equation}
  \label{qan}
  q_\alpha(x) = \fr{1}{2}\, \mathrm{Tr}\, \big{\{}\, \Gamma\,\Gamma_\alpha 
       \big{[}\, \ca{H}\,\frac{1}{-\tilde{\ca{H}}\ca{H}} - (\mu)\,\big{]}\big{\}}|_{y=x}\, ,
\end{equation}
where the operators involved are defined in (\ref{HH}) and the last term indicates the subtraction of the
same expression with $M\rightarrow\mu$. In the following we will first look at the expressions for $M$. 
Up to exponentially suppressed terms $\ca{L}\ca{L}^\dagger$ and $\ca{L}^\dagger\ca{L}$ are identical 
at the boundary of the discs (\ref{discbound}) and therefore the inverse operator in (\ref{qan}) 
considerably simplifies close to the boundary,
\begin{equation}
  \label{HHbd}
    \frac{1}{-\tilde{\ca{H}}\ca{H}}\ \approx\  
       \unit_4 \otimes \frac{1}{-\del_\alpha^2 +\mathfrak{m}^2_M -b_\alpha\del_\alpha}\, ,
\end{equation}
where we have again introduced a mass matrix $\mathfrak{m}^2_M$, which is the same as in (\ref{mMat})
but $\mu$ is replaced with $M$, as indicated by the index. At the boundary $b_\alpha$ vanishes like 
$\sim r^{-1}$, see (\ref{discbound}) and it is sufficient to know the Green's function (\ref{HHbd})
up to orders $O(r^{-2})$ to perform the integral over $\del\mathfrak{D}_{\pm}$. 

The expansion of (\ref{HHbd}) is of the form (\ref{anex}) with $\Delta_\mu$ replaced by 
$\Delta_{\mathfrak{m}_M}$ and the insertion operators are $\ca{O}(z) = b_\alpha(z)\,\del_{z^\alpha}$. 
Changing the integration variables as $z_i \rightarrow x+z_i$ the insertion becomes 
$ b_\alpha(x)\,\del_{z^\alpha}+O(r^{-2})$ and the Green's function (\ref{HHbd}) is approximated
by
\begin{equation}
  \label{greenapp}
  \frac{1}{-\del_\alpha^2 +\mathfrak{m}^2_M -b_\alpha\del_\alpha} = 
   \Delta_{\mathfrak{m}_M}(x-y) + b_\alpha(x)\, \del_{x^\alpha}\, \Delta^2_{\mathfrak{m}_M}(x-y) +O(r^{-2})\, ,
\end{equation}
where, as in the case of the disc anomaly,  we have used that the gauge fields and thus 
$b_\alpha$ (\ref{terms}) commute with the mass-matrix $\mathfrak{m}_M^2$ and the propagator 
$\Delta_{\mathfrak{m}_M}$, see below (\ref{anres}).

We note that the product $\Gamma\Gamma_\alpha$ in (\ref{qan}) is block off-diagonal, see (\ref{Htilde}), 
(\ref{gstar2}), whereas $\frac{1}{-\tilde{\ca{H}}\ca{H}}$ is block diagonal. Consequently, only the 
off-diagonal terms of $\ca{H}$ contribute in the trace (\ref{qan}), which are independent of $M$. We thus have
\begin{align}
  \label{q2}
     q_\alpha(x) =&\ \fr{1}{2}\, \mathrm{Tr}\, \big{\{}\, \Gamma\,\Gamma_\alpha \ca{H}\ \unit_4\otimes
            \Big{(}  \left[\Delta_{\mathfrak{m}_M}(x-y) -  \Delta_{\mathfrak{m}_\mu}(x-y)\right]\nonumber\\
         &\quad  + b_\alpha(x)\,\del_{x^\alpha} \left[  \Delta^2_{\mathfrak{m}_M}(x-y) -  
                           \Delta^2_{\mathfrak{m}_\mu}(x-y)\right]
           \Big{)}\,\big{\}}|_{y=x}
          +O(r^{-2})
\end{align}
The expressions in the two square brackets are regulated and well defined at $y=x$.
We also note that $\Gamma\sim\frac{\ca{M}}{\ca{M}^2 + M_P^2}$, which is an important convergence factor 
for the $p$-integration (\ref{sst}) and as in the case for the anomaly projects out possible poles for $M=0$,
which appear here also in the propagator because $I'$ in the mass-matrices of the form (\ref{mMat}) does 
not have full 
rank, see below (\ref{opsdisc}).
We can therefore safely set $y=x$ and the resulting current is then, 
\begin{align}
  \label{q1}
   q_\alpha(x) =&\ -\frac{1}{8\pi} \mathrm{Tr}\, \big{\{}\, \Gamma\Gamma_\alpha K(x)\, \unit_4\otimes 
         \log \left[\frac{\mathfrak{m}^2_M}{\mathfrak{m}^2_\mu} \right]\big{\}} \nonumber\\
         &\hspace{2.5cm}   +\frac{1}{16\pi}  \mathrm{Tr}\, \big{\{}\, \Gamma\Gamma_\alpha\Gamma_\beta\, 
           \unit_4\otimes b_\beta(x) \log \left[\frac{\mathfrak{m}^2_M}{\mathfrak{m}^2_\mu} \right]\big{\}}
          + O(r^{-2})\,.
\end{align}
For the first term we used that the derivative part of 
$\ca{H}$ (\ref{Htilde}) vanishes by symmetric integration. Similarly, single derivatives vanish for the 
second term and 
$\del_\alpha\del_\beta \Delta^2_{\mathrm{reg}} =\frac{1}{2}\,\delta_{\alpha\beta}\,\del^2  \Delta^2_{\mathrm{reg}}$. 

Seemingly the limit $\mu\rightarrow\infty$ of (\ref{q1}) does not exist, but 
this is the last step to carry out in (\ref{sst}) and we note that after the $p$-integration the 
limit does exist. However, before doing so we use that $\unit_4\otimes b_\alpha = \{\Gamma_\beta,K\}$ 
and that the matrix  $\log \left[\mathfrak{m}^2_M / \mathfrak{m}^2_\mu \right]$ commutes 
with $\Gamma_\alpha$ such that with the relations (\ref{gstar}) one can show that the two terms in 
(\ref{q1}) in fact cancel. Thus $q_\alpha(x) =0$ and the disc surface term does not contribute,
as stated in the main text.

\section{Cartan-Weyl basis for $U(N)$}\label{CW}

We give some basic definitions for the Lie-algebra of $U(N)$ and thereby introduce the 
conventions that are used in the main text. Throughout we use the notation $\bs{e}_{m=1,\ldots,N}$ 
for the canonical basis of the vector space $\mathbb{R}^N$, i.e. with unit entry at position $m$ and otherwise
zero (in components $e^i_{m} = \delta^i_m$)

In addition to the $\mathrm{rank}=N-1$ 
mutually commuting generators of $\mathfrak{su}(N)$ one has the unit matrix (\ref{eq:un})
in the Cartan subalgebra $\mathfrak{h} \subset \mathfrak{u}(N)$. The absence of the traceless-condition allows 
the choice of a rather convenient basis for $\mathfrak{h}$. Together with the ladder 
operators these form a basis of $\mathfrak{u}(N)$:
\begin{align}
  \label{gens}
  H_i =  \frac{1}{\sqrt{2}}\begin{bmatrix}\ddots& &\\[-8pt] &1 &\\[-9pt] & &\ddots\end{bmatrix} 
   \quad , \quad\quad 
   E_{\bs{\alpha}_{mn}} = \frac{1}{\sqrt{2}}\begin{bmatrix}  & : & &\\[-3pt]
      \cdots& 1&\cdot\cdot  &\\ & \vdots& &   \end{bmatrix}\ .
\end{align}
The $N$ Cartan generators $H_i$ have the non-vanishing entry at the $i$'th position
on the diagonal, whereas the $N(N-1)$ ladder operators $E_{\bs{\alpha}_{mn}}$ have the only 
non-vanishing entry at (row,column) = $(m,n)$ with $m\neq n$.
The ladder operators are labeled by the $N$-component roots 
$\bs{\alpha}_{mn} = (\alpha^i_{mn})$, which are defined by the second of the following commutation
relations:\footnote{Here and in the following we 
 often omit the labels $m,n$ for the different roots. A root symbol  $\bs{\alpha},\bs{\beta}$ etc. 
stands then for a particular root, i.e. particular values for $m$, $n$.}
\begin{align}
  \label{slgebra}
   [\,H_i\, ,H_j\,]=0\, , \ \ \ [\,H_i\,, E_{\bs{\alpha}}\,] = \alpha^i\,E_{\bs{\alpha}}\, ,\ \ \  
        [\,E_{\bs{\alpha}}\, , E_{-\bs{\alpha}}\,] = \alpha^i\,H_i\, .
\end{align}
The $N(N-1)$ roots are vectors $\in \mathbb{R}^N$ and are of the form
\begin{equation}
  \label{roots}
    \bs{\alpha}_{mn} = -  \bs{\alpha}_{nm}  = \frac{1}{\sqrt{2}} (\bs{e}_m -\bs{e}_n) 
   \quad \Rightarrow\quad \bs{\alpha}^2=1\, .
\end{equation}
The set of roots $\Sigma:=\{\bs{\alpha}_{mn}\}$ can be divided into positive and negative roots,
$\Sigma = \Sigma^+ \cup\, \Sigma^-$ by the 
convention that a root is positive if the first non-vanishing entry is positive and otherwise negative. 
The positive 
roots are therefore given by $\Sigma^+=\{\bs{\alpha}_{m<n}\}$. The negative roots are just the negative thereof.
Also the positive roots are not linearly independent. A set of linear independent roots that 
generates all roots by linear combinations with integer coefficients of a common sign is called simple roots:
\begin{equation}
  \label{simple}
  \bs{\beta}_{a}=\frac{1}{\sqrt{2}}(\bs{e}_a - \bs{e}_{a+1})\, ,\ \ a=1,\ldots,N-1
  \qquad \Rightarrow\quad \Sigma^\pm \ni \bs{\alpha}_{mn} = \pm \sum_{a=m}^{n-1} \bs{\beta}_{a}\, ,
\end{equation}
thus positive/negative roots have positive/negative integer coefficients.

From the definition of the ladder operators and the roots follows that $E_{\bs{\alpha}}^\dagger = E_{-\bs{\alpha}}$. 
By defining the hermitian generators 
\begin{equation}
  \label{hermbas}
 T^{(+)}_{\bs{\alpha}} =\frac{1}{\sqrt{2}} ( E_{\bs{\alpha}} + E_{-\bs{\alpha}})\, ,\qquad
 T^{(-)}_{\bs{\alpha}} =\frac{i}{\sqrt{2}} ( E_{\bs{\alpha}} - E_{-\bs{\alpha}})\, ,
\end{equation}
a hermitian basis of generators with the normalization 
as given in (\ref{eq:un}) is then given by 
$\{\,H_i\,, T^{(+)}_{\bs{\alpha}}\,, T^{(-)}_{\bs{\alpha}}\,|\,\bs{\alpha}\in \Sigma^+\, \}$. 
These relations are completely standard and are the same as for $\mathfrak{su}(N)$ except for the fact that 
the Cartan subalgebra has now one more element. Basically the only choice of conventions that matters is the 
normalization (\ref{roots}). 

It will be convenient to have an explicit matrix realization of the Cartan generators in the adjoint 
representation. In the  hermitian basis introduced  here it is given by
\begin{equation}
  \label{Had}
   H_k^{\mathrm{ad}} = \begin{bmatrix} 0 & \cdot\cdot \\ : & 0 & i A_k\\
                           & - iA_k& 0\end{bmatrix}\, , 
\end{equation}
where $A_k$ is the diagonal matrix of the $k$'th component of all positive roots, i.e. 
$A_k = \mathrm{diag}(\alpha^k_{1,2},\ldots, \alpha^k_{N-1,N})$. This implies the normalization
\begin{equation}
  \label{adnorm}
  \mathrm{Tr}_{\mathrm{ad}}\{H_iH_j\} = 
   2 \sum_{\bs{\alpha}_{\mathrm{pos}}} {\alpha}^i\, {\alpha}^j 
     =\sum_{\bs{\alpha}} {\alpha}^i\, {\alpha}^j = N\, \delta^{ij}\, .
\end{equation}
Note, in the 
non-hermitian basis (\ref{gens}) the adjoint matrix of $H_i$ is diagonal.

%%%%%%%%%%% R E F E R E N C E S %%%%%%%%%%%%%%%%%

\bibliographystyle{JHEP}
%\bibliography{refs2}

\providecommand{\href}[2]{#2}\begingroup\raggedright\endgroup

\end{document}